\documentclass[twocolumn]{aastex63}
\usepackage{amssymb}
\usepackage{amsmath}

\def\cm{{\rm\thinspace cm}}

\def\erg{{\rm\thinspace erg}}
\def\eV{{\rm\thinspace eV}}
\def\g{{\rm\thinspace g}}

\def\K{{\rm\thinspace K}}
\def\keV{{\rm\thinspace keV}}
\def\km{{\rm\thinspace km}}

\def\m{{\rm\thinspace m}}
\def\Mpc{{\rm\thinspace Mpc}}

\def\s{{\rm\thinspace s}}

\def\ergpcmsqps{\hbox{$\erg\cm^{-2}\s^{-1}\,$}}

\def\ergps{\hbox{$\erg\s^{-1}\,$}}

\def\kmps{\hbox{$\km\s^{-1}\,$}}

\def\pcmsq{\hbox{$\cm^{-2}\,$}}

\def\ps{\hbox{$\s^{-1}\,$}}

\def\kmpspMpc{\hbox{$\kmps\Mpc^{-1}$}}

\def\h18{\hbox{H1821$+$643\,}}

\received{-}
\revised{-}
\accepted{-}
\submitjournal{ApJ}

\shorttitle{Astrophysical limits on very light ALPs}
\shortauthors{C.S. Reynolds et al.}

\begin{document}

\title{Astrophysical limits on very light axion-like particles from Chandra grating spectroscopy of NGC~1275}

\correspondingauthor{Christopher S. Reynolds}
\email{csr12@ast.cam.ac.uk}

\author[0000-0002-1510-4860]{Christopher S. Reynolds}
\affiliation{Institute of Astronomy, University of Cambridge, Madingley Road, Cambridge CB3~OHA, UK}

\author[0000-0001-7271-4115]{M.~C.~David Marsh}
\affiliation{The Oskar Klein Centre, Department of Physics, Stockholm University, Stockholm 106 91, Sweden}

\author{Helen~R.Russell}
\affiliation{School of Physics and Astronomy, University of Nottingham, Nottingham NG7 2RD, UK}
\affiliation{Institute of Astronomy, University of Cambridge, Madingley Road, Cambridge CB3~OHA, UK}

\author[0000-0002-9378-4072]{Andrew C. Fabian}
\affiliation{Institute of Astronomy, University of Cambridge, Madingley Road, Cambridge CB3~OHA, UK}

\author{Robyn Smith}
\affiliation{Dept. of Astronomy, University of Maryland, College Park, MD~20742, USA}

\author[0000-0002-6562-8654]{Francesco Tombesi}
\affiliation{Dept/ of Physics, University of Rome ``Tor Vergata'', Via della Ricerca Scientifica 1, I-00133 Rome, Italy}
\affiliation{Dept. of Astronomy, University of Maryland, College Park, MD~20742, USA}
\affiliation{NASA/Goddard Space Flight Center, Code 662, Greenbelt, MD 20771, USA}?\affiliation{INAF Astronomical Observatory of Rome, Via Frascati 33, 00078 Monteporzio Catone (Rome), Italy}

\author[0000-0002-3158-6820]{Sylvain Veilleux}
\affiliation{Dept. of Astronomy, University of Maryland, College Park, MD~20742, USA}
\affiliation{Institute of Astronomy, University of Cambridge, Madingley Road, Cambridge CB3~OHA, UK}
\affiliation{Kavli Institute for Cosmology, University of Cambridge, Madingley Road, Cambridge CB3~OHA, UK}

\begin{abstract}
\noindent Axions/axion-like particles (ALPs) are a well motivated extension of the Standard Model and are generic within String Theory.  The X-ray transparency of the intracluster medium (ICM) in galaxy clusters is a powerful probe of light ALPs (with mass $<10^{-11}\eV$); as X-ray photons from an embedded or background source propagate through the magnetized ICM, they may undergo energy-dependent quantum mechanical conversion into ALPs (and vice versa), imprinting distortions on the X-ray spectrum.  We present {\it Chandra} data for the active galactic nucleus NGC~1275 at the center of the Perseus cluster. Employing a 490\,ks High-Energy Transmission Gratings (HETG) exposure, we obtain a high-quality 1--9\,keV spectrum free from photon pileup and ICM contamination.  Apart from iron-band features, the spectrum is described by a power-law continuum, with any spectral distortions at the $<3\%$ level. We compute photon survival probabilities as a function of ALP mass $m_a$ and ALP-photon coupling constant $g_{a\gamma}$ for an ensemble of ICM magnetic field models, and then use the NGC~1275 spectrum to constraint the $(m_a, g_{a\gamma})$-plane.  Marginalizing over magnetic field realizations, the 99.7\% credible region limits the ALP-photon coupling to $g_{a\gamma}<6-8\times 10^{-13}\, {\rm GeV}^{-1}$ (depending upon magnetic field model) for masses $m_a<1\times 10^{-12}\eV$.  These are the most stringent limit to date on $g_{a\gamma}$ for these light ALPs, and have already reached the sensitivity limits of next-generation helioscopes and light-shining-through-wall experiments.  We highlight the potential of these studies with the next-generation X-ray observatories {\it Athena} and {\it Lynx}, but note the critical importance of advances in relative calibration of these future X-ray spectrometers.
\vspace{1cm}
\end{abstract}





\section{Introduction}\label{intro}

\noindent Astrophysical observations have great potential to uncover new physics beyond the Standard Model (SM). Indeed, the clearest experimental indications that new physics must be manifest at low energies is the astrophysical observation that SM particles and fields only account for 4\% of the energy density of our Universe \citep{ade:14a}.

Especially interesting, and the focus of this paper, are axion-like particles \citep{graham:15a,irastorza:18a}. The axion is a consequence of a well-motivated extension to the SM, namely the Peccei-Quinn (PQ) mechanism that protects the strong interaction from CP (i.e.~time-reversal) violating effects \citep{peccei:77a,weinberg:78a,wilczek:78a}.  This Quantum Chromodynamics (QCD) axion couples to two photons with a coupling strength $g_{a\gamma}$ that is proportional to their mass (and inversely proportional to the energy scale of PQ symmetry breaking). Furthermore, many extensions of the SM, including in particular string theory,  commonly feature very light axion-like particles (ALPs) that do not couple to the strong interactions, but can interact with photons with a strength that is independent of their masses. The coupling constant $g_{a\gamma}$ and ALP mass $m_a$ can, in a natural way, be small compared with other SM mass scales.  
%
%
Axions and ALPs may be produced in the early Universe via non-thermal mechanisms, vacuum realignment and the decay of topological defects, and hence would be produced with very small velocity dispersion. Thus, despite being very light, they can still be a viable candidate for cold dark matter \citep{Preskill:1982cy,abbott:83a,dine:83a}.

{ There is a rich literature on searches for and constraints on axions/ALPs using astrophysical and cosmological observations; for an up-to-date review see \cite{tanabashi:18a}. The historically most important limit  on sufficiently light ALPs comes from SN1987A \citep{brockway:96a,grifols:96a,payez:15a}. Other searches include those based on the structure and luminosity of stars \citep{vysotsskii:78a}, the polarization of the cosmic microwave background \citep{tiwari:12a,mukherjee:19a}, the timing of radio pulsars and fast radio bursts \citep{caputo:19a}, and the possible discrepancies between the excess of Cosmic IR background radiation at 1$\mu$m and the TeV opacity of the Universe \citep{kohri:17a}. }

The transparency of astrophysical systems is a particularly simple and effective way to search for the effects of ALPs; as astrophysical photons traverse through cosmic magnetic fields, they are susceptible to conversion to an ALP via the two-photon interaction described by the Lagrangian term
\begin{equation}
{\cal L}_a=-g_{a\gamma}a{\bf E}\cdot{\bf B}.
\end{equation}
where $a$ is the ALP field, and ${\bf E}$ and ${\bf B}$ are the electric and magnetic fields. The transparency of astrophysical systems can be used to set upper limits on $g_{a\gamma}$ and may be a route to the eventual detection of ALPs.  

X-ray observations of active galactic nuclei (AGN) in rich clusters of galaxies are particularly suited to ALP searches. Faraday rotation measure (RM) studies demonstrate that the hot intracluster medium (ICM) in such clusters is magnetized, with a ratio of thermal-to-magnetic pressure of $\beta\sim 100$ \citep{taylor:06a}. Furthermore, we expect significant regions of this field to be coherent on scales $1-10$\,kpc \citep{vacca:12a}. The result is that, if ALPs exist with a sufficiently high coupling to photons, clusters will be efficient converters of X-ray photons into ALPs. For typical parameters relevant to the ICM, the conversion probability will be energy dependent thereby imprinting distortions into the observed spectrum of any embedded (or background) object, with a precise form that depends upon the magnetic field structure as well as the ALP properties.  We note that these transparency studies probe the physics of ALPs independently of whether they actually constitute a significant component of the non-baryonic dark matter.

\cite{wouters:13a} used {\it Chandra} imaging spectroscopy of the central AGN Hydra-A in the Hydra cluster of galaxies to set an upper limit on spectral distortions from ALPs, showing that any ALP with mass $m_a<7\times 10^{-12}\,{\rm eV}$ must have $g_{a\gamma}<8.3\times 10^{-12}\,{\rm GeV}^{-1}$ (95\% confidence level).  Subsequently, \cite{berg:17a} examined the AGN NGC~1275 at the center of the Perseus cluster, the target of our current study, using {\it Chandra} and {\it XMM-Newton} imaging spectroscopy.  NGC~1275 is almost 100$\times$ brighter than Hydra-A in the 2--10\,keV band and, furthermore, does not have the heavy intrinsic absorption of Hydra-A. While this dramatically reduces the statistical errors on the spectrum of the NGC~1275, the high source flux creates systematic issues.  For the large body of {\it Chandra} imaging data (almost 1Ms) for which NGC~1275 is close the optimal aim-point, the sub-arcsecond focusing leads to very severe photon-pileup; thus there are very strong spectral distortions that are entirely instrumental in origin. For this reason, \cite{berg:17a} focus their attention on the smaller quantities of {\it Chandra} data for which NGC~1275 is off-axis and on data from {\it XMM-Newton}. Since the poorer focus for these datasets reduces but does not eliminate the effect of pileup, they employ the pileup model of \cite{davis:01a} and conclude that any spectral distortions must be below the 10\% (once smoothed to the spectral resolution of the CCD detectors in {\it Chandra} and {\it XMM-Newton}). The resulting ALP constraint is $g_{a\gamma}<4\times 10^{-12}\,{\rm GeV}^{-1}$ for massless ALPs.
 \cite{Chen:2017mjf} extended this analysis to massive ALPs, and verified that this limit applies to $m_a\lesssim 10^{-12}\eV$.\footnote{For ALP limits from Chandra observations of seven sources (with more poorly constrained magnetic fields), see also \cite{Conlon:2017qcw}.}

To circumvent the photon pileup issue, \cite{marsh:17a} used 370\,ks of short (0.4\s) frame time {\it Chandra} imaging spectroscopy of the core of M87 in the Virgo cluster.  Combined with the fact that this AGN is almost an order of magnitude fainter than NGC1275, the short frame time exposures led to a high-quality spectrum with negligible pileup, and ALP constraints of $g_{a\gamma}<2.6\times 10^{-12}\,{\rm GeV}^{-1}$ for $m_a< 10^{-12}\eV$ (95\% confidence level).

In this paper, we present new {\it Chandra} observations of NGC~1275 that set the tightest limits to date on light ALP conversion. By employing the High-Energy Transmission Gratings (HETG) and investing almost 500\,ks of on-source exposure, we obtain a high-quality separation of the AGN emission from the ICM with no discernible photon-pileup. Outside of the astrophysically-rich  iron-band (6--7\,keV), we find that the resulting 1--9\,keV spectrum of the AGN is well described by a power-law continuum (modified only by the effects of absorption by cold gas in our Galaxy) with any remaining spectral distortions below the 3\% level. 

We proceed to marginalize over a set of representative realisations of the magnetic field  to determine the posterior distribution  on the $(m_a, g_{a\gamma})$-plane.  The resulting 99.7\% credible region limits the ALP-photon coupling to $g_{a\gamma}<6-8\times 10^{-13}\, {\rm GeV}^{-1}$ (depending upon the magnetic field model) for most masses $m_a<1\times 10^{-12}\eV$.  This is the most stringent limit to date on the ALP-photon coupling.

For one of our magnetic field models, the posterior peaks at a non-vanishing value of $g_{a \gamma}$, and the the 95\% credible region pick out a preferred non-zero value of the couple constant, $g_{a\gamma}\approx 4-8\times 10^{-13}\, {\rm GeV}^{-1}$, for a range of masses $m_a<1\times 10^{-12}$.  We attribute this ``detection'' to residual, low-level, errors in the {\it Chandra}/HETG calibration.


This paper is organized as follows. Section~\ref{reduction} presents the new observations and describes some subtleties encountered during the data reduction. After discussing the modeling of the ICM magnetic field and the associated ALP distortions in Section~\ref{alpmodels}, the new constraints on ALPs are given in Section~\ref{results}. We put these results into context and draw our conclusions in Section~\ref{conclusions}.  When necessary, we assume a standard {\it Planck} cosmology with $H_0=68\kmpspMpc$ \citep{ade:14a} which, at a redshift of $z=0.0173$ \citep{hitomi:18a}, places NGC~1275 at a distance of 76\,Mpc.

\section{Observations, Data Reduction and Initial Spectral Fitting}\label{reduction}

\begin{figure*}[t]
\begin{center}
\includegraphics[width=0.9\textwidth]{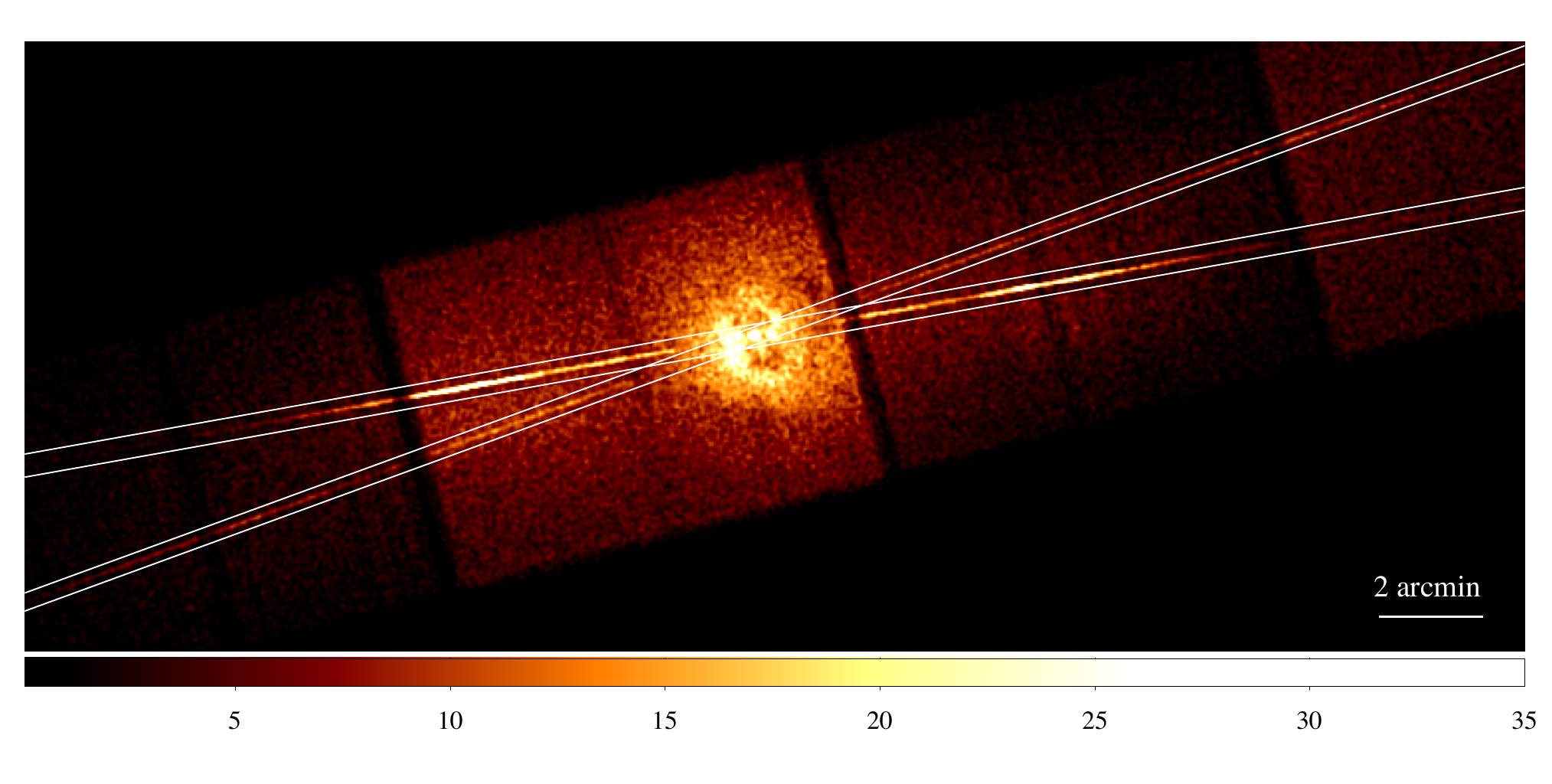}
\end{center}
\caption{Full-band image of the ACIS-S array for ObsID~20449 { (exposure 45\,ks). For purposes of display only, the raw pixel data have been binned by $4\times 4$, and the color bar shows total photon count per (new) binned pixel for this ObsID.  The extraction region for the dispersion spectrum is shown in white.}}
\label{fig:hetgimage}
\vspace{0.5cm}
\end{figure*}

\noindent {\it Chandra} observed NGC~1275 in 15 separate segments (ObsIDs) between 24-Oct-2017 and 5-Dec-2017 using the HETG read out on the Advanced CCD Imaging Spectrometer (ACIS) S array. The total on-source exposure time was 490\,ks. To ameliorate the risk of modest photon pileup in the event that the source was brighter than expected, we turned off the two outlying ACIS-S chips and used 1/2 sub-arrays on the remaining four chips, resulting in a reduction of the frame readout time to 2.4s with no loss of observing efficiency. While loss of the two outlying chips in principle affects our ability to observe the softest regions of the spectrum, the contaminant that has built up on the ACIS optical blocking filters unavoidably removes those soft photons anyway. 

Figure~\ref{fig:hetgimage} shows the image of the ACIS-S array for one of our ObsIDs (20449).  This is dominated by zeroth order image of core of the Perseus cluster with its famous cavity system \citep{fabian:00a}. The two two-sided dispersed spectra of the bright central AGN emission, one from the High Energy Grating (HEG) and one from the Medium Energy Grating (MEG), are clearly visible. Although the gratings are slitless and so the cluster light is also dispersed, it is clear that the AGN light is distinct and can be well isolated from the bulk of the ICM emission. Order-sorting, whereby only photons with CCD-detected energies compatible with their spatial position along the dispersion spectrum are accepted,  enables further isolation of the AGN spectrum.

Much but not all of our data reduction is standard. All data were reprocessed with CIAO-4.10 and CALDBv4.8.1. The extraction of the AGN spectra for each ObsIDs then follows the standard CXC science threads\footnote{http://cxc.cfa.harvard.edu/ciao/threads/spectra\_hetgacis/} with two exceptions. Firstly, we halve the width of the extraction regions ({\tt width\_factor\_hetg=18}) in order to  reduce MEG/HEG overlap at the centre of the array and hence access the higher energy band in the HEG. Secondly, the standard algorithm for automatically locating the zeroth order image and hence setting the energy scale of the spectrum failed for most of the ObsIDs, presumably due to the surrounding high-surface brightness and structured ICM. Instead, we force the zeroth order point to be at the known coordinates of NGC1275, and visually confirm for each ObsID that this correctly locates the zeroth order image of the point-like AGN (i.e.~that astrometry errors are within one pixel or 0.5\,arcsec). The result is four spectra and associated background spectra, response matrices and effective area files for each ObsID, namely the $+1$ and $-1$ order spectra for each of the MEG and HEG. In a final step, we combine the spectra to produce a single HEG and a single MEG spectrum (with associated background spectra, response matrices, and effective area files), summing the $+/-1$ orders from all ObsIDs.

All spectral fitting presented in this paper uses the $1-7\keV$ band for the MEG, the $1.5-9\keV$ band for the HEG, and employs Cash ($C$) statistic minimization to fit the unbinned spectrum, modified to allow for the subtraction of a background spectrum\footnote{See https://heasarc.gsfc.nasa.gov/xanadu/xspec/manual/ node304.html\#AppendixStatistics}. Fitting is performed with the XSPECv12.10.1 code \citep{arnaud:96a}. 

\begin{figure}[t]
\includegraphics[width=0.45\textwidth]{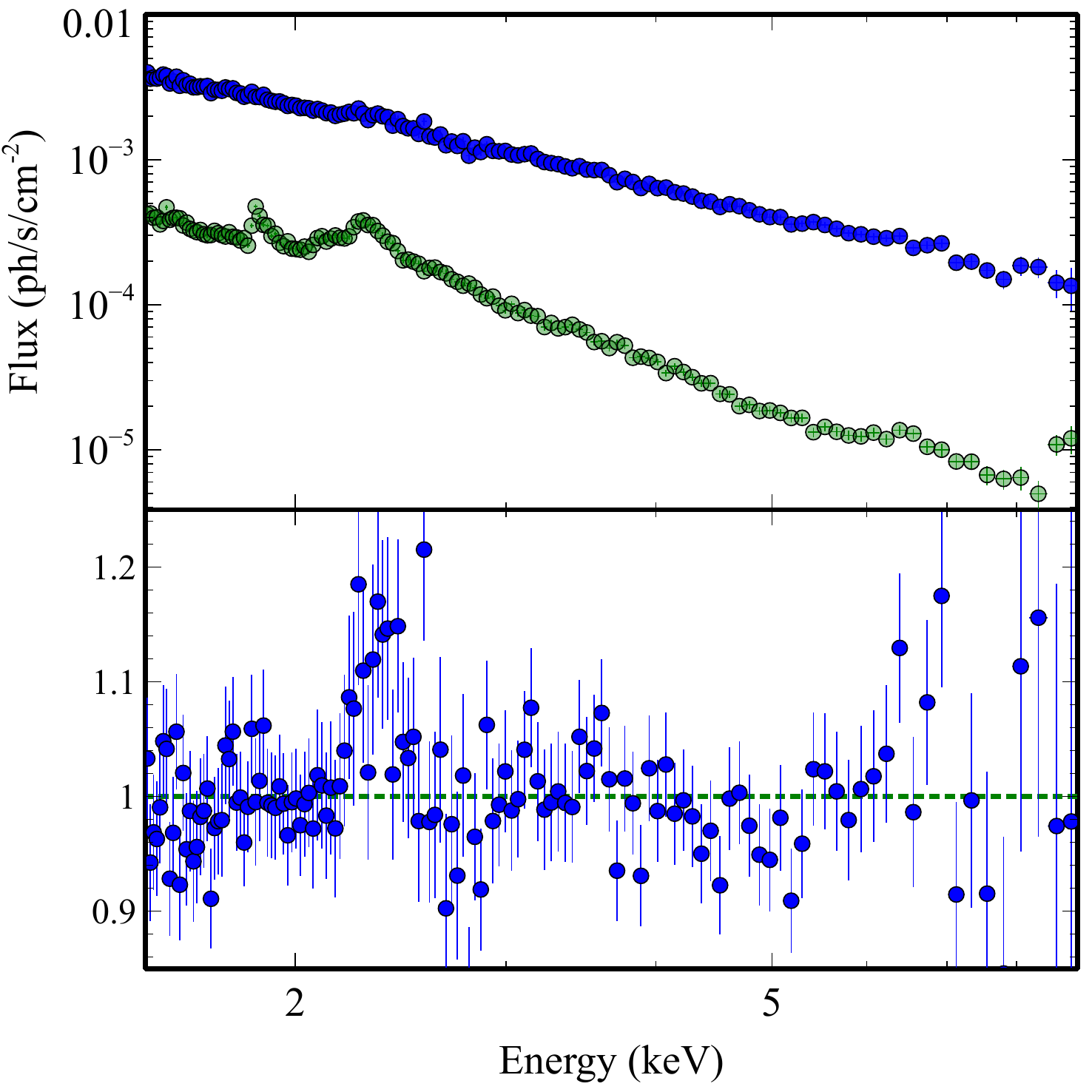}
\label{fig:heg_rawback}
\caption{{ An illustration of the background under-subtraction issue when using the default-normalized background spectra.  Shown here is the best-fitting power-law model to the combined HEG spectrum (blue) that has been background subtracted using the background-spectrum obtained from the standard pipeline (green). The significant deviations from the power-law at $\sim 2.5\keV$ clearly mirror a feature in the background spectrum.}}
\end{figure}

An initial fit of the spectra with a power-law continuum modified by Galactic absorption \citep[$N_H=1.32\times 10^{21}\pcmsq$; ][]{kalberla:05a} finds 10--15\% excesses in three energy bands; below 1.3\,keV (MEG only), $2.2-2.5\keV$ (HEG and MEG), and $6-7\keV$ (principally in the HEG). The $6-7\keV$ structure corresponds to the well-established iron fluorescent line from cold gas in the vicinity of the AGN \citep{hitomi:18a} and will be the subject of another paper (Reynolds et al., in prep). { As illustrated in Fig.~\ref{fig:heg_rawback} for the HEG,} the other residuals closely mirror structure in the background spectra (which is actually dominated by the core ICM emission) and suggests that the standard grating extraction algorithms have underestimated the background normalization.  { Broadly, this is not a surprise. The background spectrum is determined from strips that flank the source extraction region with algorithms that are designed and optimized for a spatially-uniform background around a point source.  In our case, the ICM emission that forms our background is centrally concentrated around the AGN, so we expect that the spectrum extracted from the background/flanking regions will be normalized too low. It is not possible, however, to simply estimate the size of this effect from the surface brightness profile --- the dispersion of the ICM emission in this slit-less grating system together with the order-sorting algorithm (where the intrinsic energy resolution of the ACIS is used to reject all photons that definitely lie outside of the expected map of dispersed position to energy) makes the background normalization a non-trivial function of the ICM spatial and spectral structure.}

{ Instead, we follow an empirical approach.  By scanning through a range of possible renormalization values, we find that the $C$-statistic of the power-law fit is minimized if the HEG and MEG backgrounds are scaled up by factors of 2.32 and 1.92 respectively.  This reduces the deviations from the power-law to below the 3--5\% level.  We validate this process with simulations.  We use the MARX package to simulate an HETG observation of a Perseus-like cluster with the following components (i) a point-like AGN with a power-law spectrum (photon index $\Gamma=1.9$), (ii) an ICM core described by a $\beta$-profile with core radius 2\,arcmin and an optically-thin thermal plasma spectrum (temperature $kT=4\keV$), and (iii) a model for the ICM cavity/shell structure consisting of two annular rings offset so that they just overlap at the AGN with inner radius 0.6\,arcmin, outer radius 1\,arcmin, and an optically-thin thermal plasma spectrum (temperature $kT=2\keV$).  We then pass the simulated events files through the standard extraction pipeline used for the real data.  This confirms that, when attempting to analyze the spectrum of the AGN, the spatial structure of the cluster leads to a systematic underestimate in the normalization of the background spectrum by approximately a factor of two.}

\begin{table*}
\caption{Power-law fits (modified by Galactic absorption) to the combined first-order HEG and MEG spectra.}
\begin{center}
\begin{tabular}{lcc}\hline\hline
Parameter & HEG value & MEG value \\\hline
Galactic absorption, $N_H$ & $1.32\times 10^{21}\pcmsq$ (fixed) & $1.32\times 10^{21}\pcmsq$ (fixed) \\
Photon index, $\Gamma$ &  $1.852\pm 0.017$  & $1.882\pm 0.011$  \\
Normalization, $A$ &  $(7.74\pm 0.14)\times 10^{-3}$  & $(8.34\pm 0.08)\times 10^{-3}$ \\
Flux (1--9\,keV) & $3.06\times 10^{-11}\ergpcmsqps$ & $3.18\times 10^{-11}\ergpcmsqps$ \\
C/DOF & 2835/2750 & 2045/2122 \\\hline
\end{tabular}
\end{center}
\label{tab:basicparams}
\end{table*}

\begin{figure*}[t]
\includegraphics[width=0.45\textwidth]{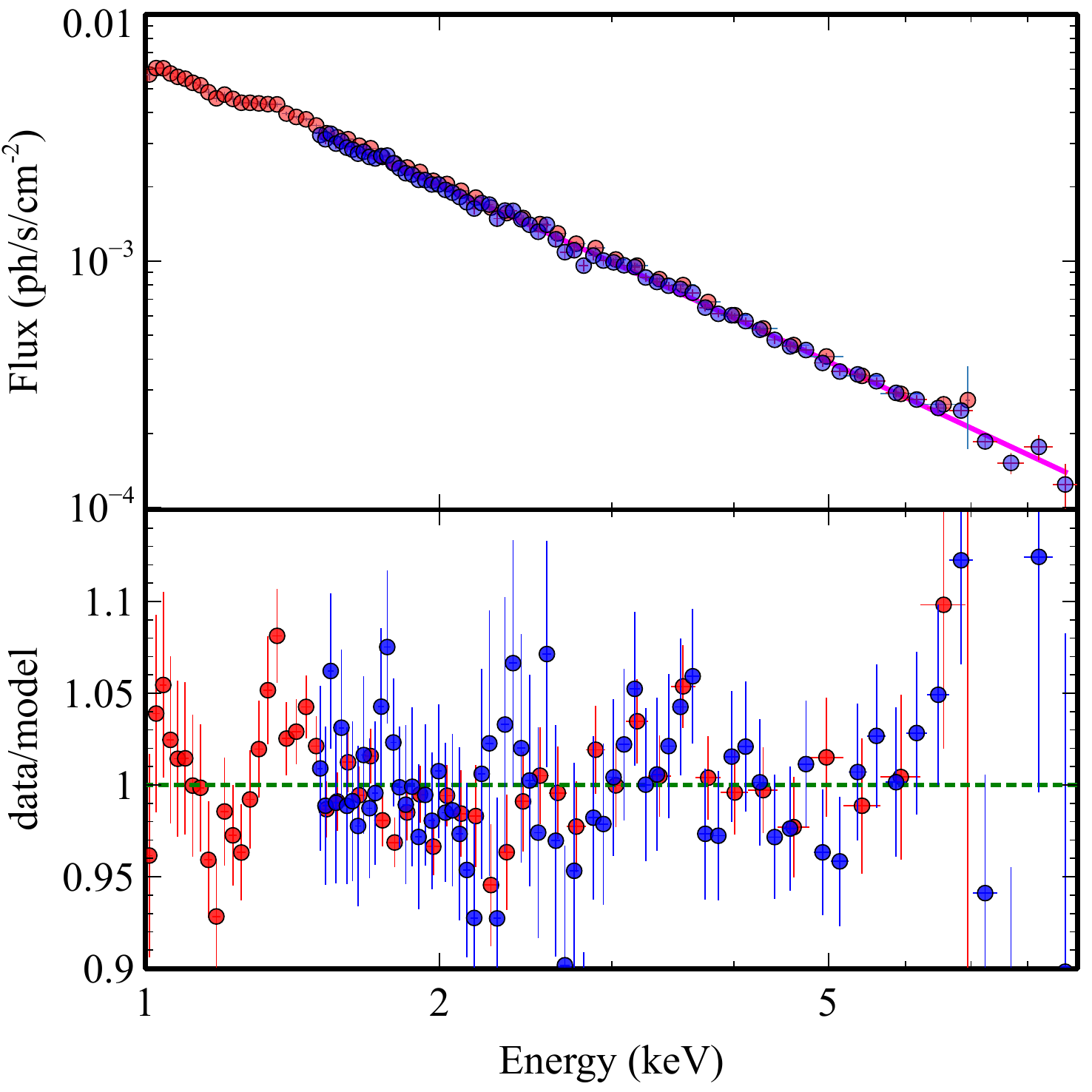}
\hspace{0.5cm}
\includegraphics[width=0.45\textwidth]{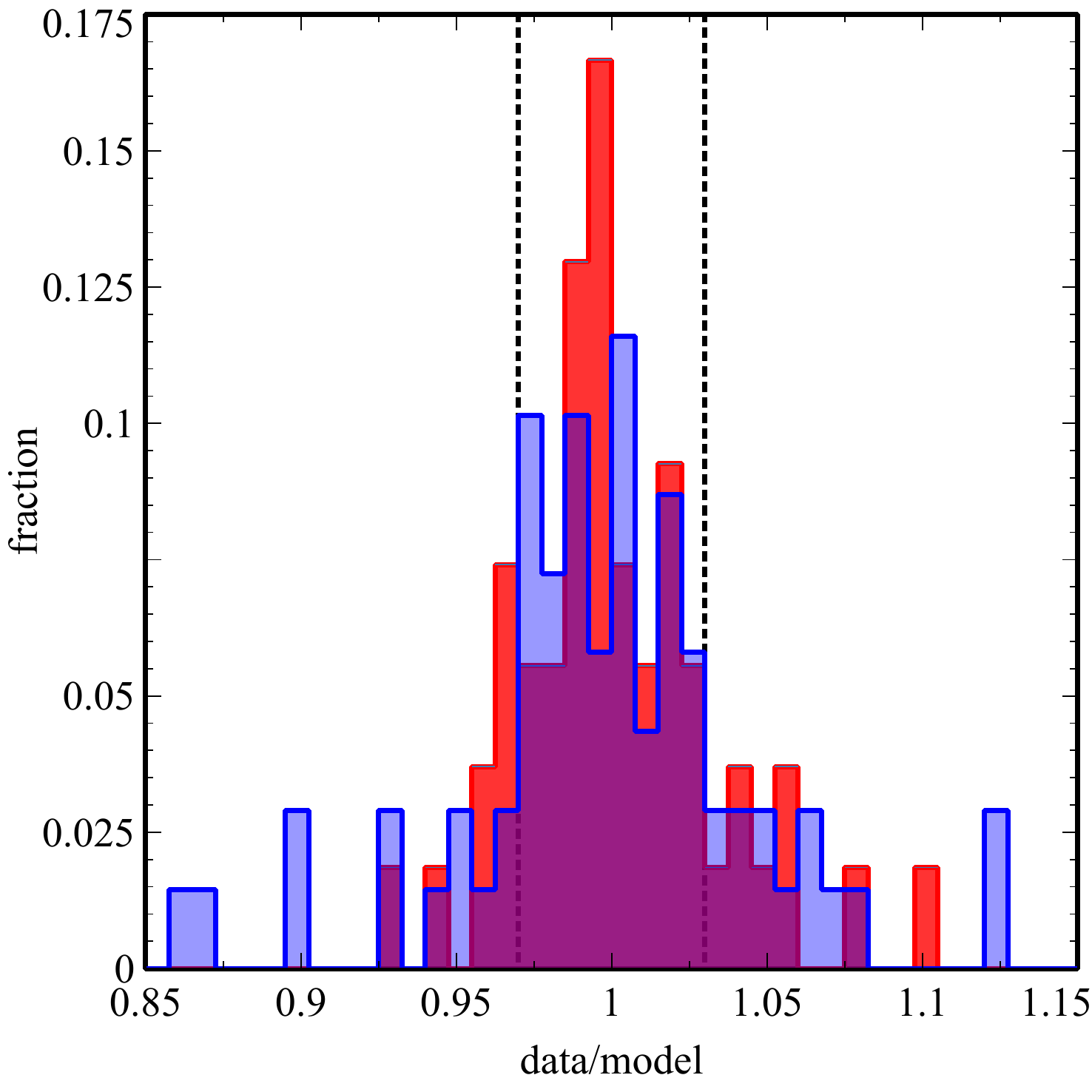}
\caption{{\it Left panel : }Best fitting power-law model to the combined HEG (blue) and MEG (red) first-order spectra (top) with corresponding ratios of the data to the best fitting model.  The data have been heavily binned for the purposes of plotting, but spectral fitting is performed on unbinned data.  {\it Right panel : }Distribution of the ratios of the data to the best fitting model for the HEG (blue) and MEG (red).  The vertical dotted lines denote the $\pm 3$\% levels.}
\label{fig:plratio}
\vspace{0.5cm}
\end{figure*}

With these adjusted background normalizations, the AGN photons comprise 80\% and 82\%  of the HEG and MEG spectra, respectively. A joint HEG/MEG fit gives a power-law index of $\Gamma=1.890\pm 0.009$  and normalization at 1keV of $A=(8.28\pm 0.06)\times 10^{-3}\,{\rm ph}\ps\keV^{-1}\pcmsq$, with $C=4956$ for 4874 degrees of freedom (DOF).  This corresponds to a 1--9\,keV band flux and luminosity of $F_{1-9\keV}=3.8\times 10^{-11}\ergpcmsqps$ and $L_{1-9\keV}=2.2\times 10^{43}\ergps$, respectively.  If we allow the power-law parameters to float freely between the MEG and HEG spectra, the fit improves significantly ($\Delta C=-76$) and we find marginally-significant slope and highly-significant normalization differences between the two gratings (Table 1). While highly-significant, the absolute flux difference is below 10\% and hence within the realm of what can reasonably be attributed to an instrumental cross-calibration uncertainty. Hence, for all subsequent fitting in this paper, we will permit normalization and slope offsets between the HEG and MEG spectra.

Figure~\ref{fig:plratio} shows this free fit of the absorbed power-law to the HEG and MEG spectra, heavily binned for plotting purposes (but unbinned for fitting purposes).  The residuals about the power-law are not entirely random, with a slight broad dip around 1.2\,keV, a very subtle broad hump between 3--4\,keV, and an obvious feature in the iron band (6.4--7\,keV).  Still, outside of the iron band, the remaining residuals are less than 5\% and mostly less than 3\% and so are entirely consistent with the expected level of residual effective area calibration errors \citep[see Fig.~7 of][]{marshall:12a}.  Having probably reached the level dominated by systematic calibration uncertainties, this is the highest quality 1-9\,keV band spectrum of this AGN obtained to date and, as we shall see, permits the most sensitive search yet for light ALPs.

\section{Modeling the ALP signatures}\label{alpmodels}

\begin{figure*}
\begin{center}
\includegraphics[width=0.4\textwidth]{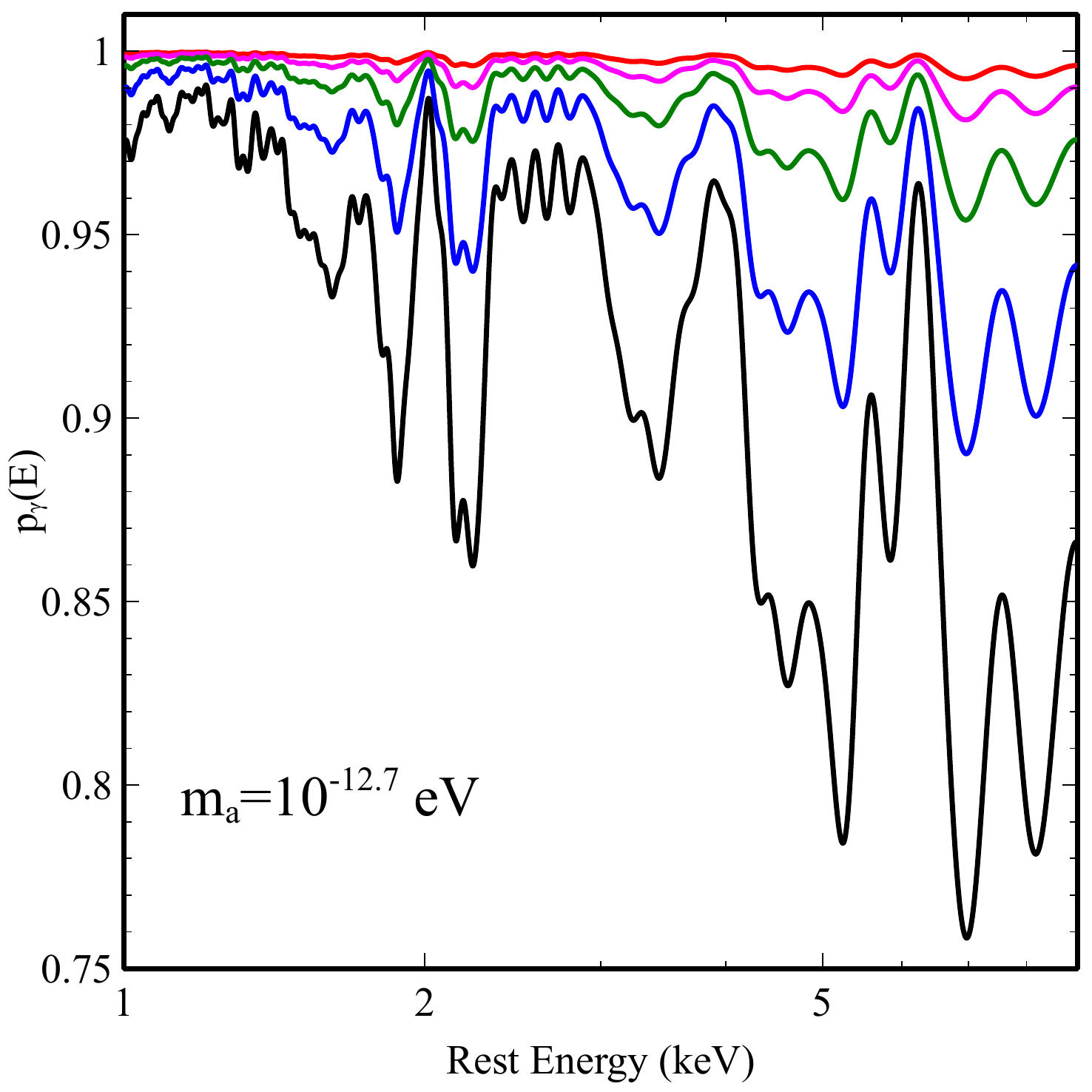}
\hspace{0.5cm}
\includegraphics[width=0.4\textwidth]{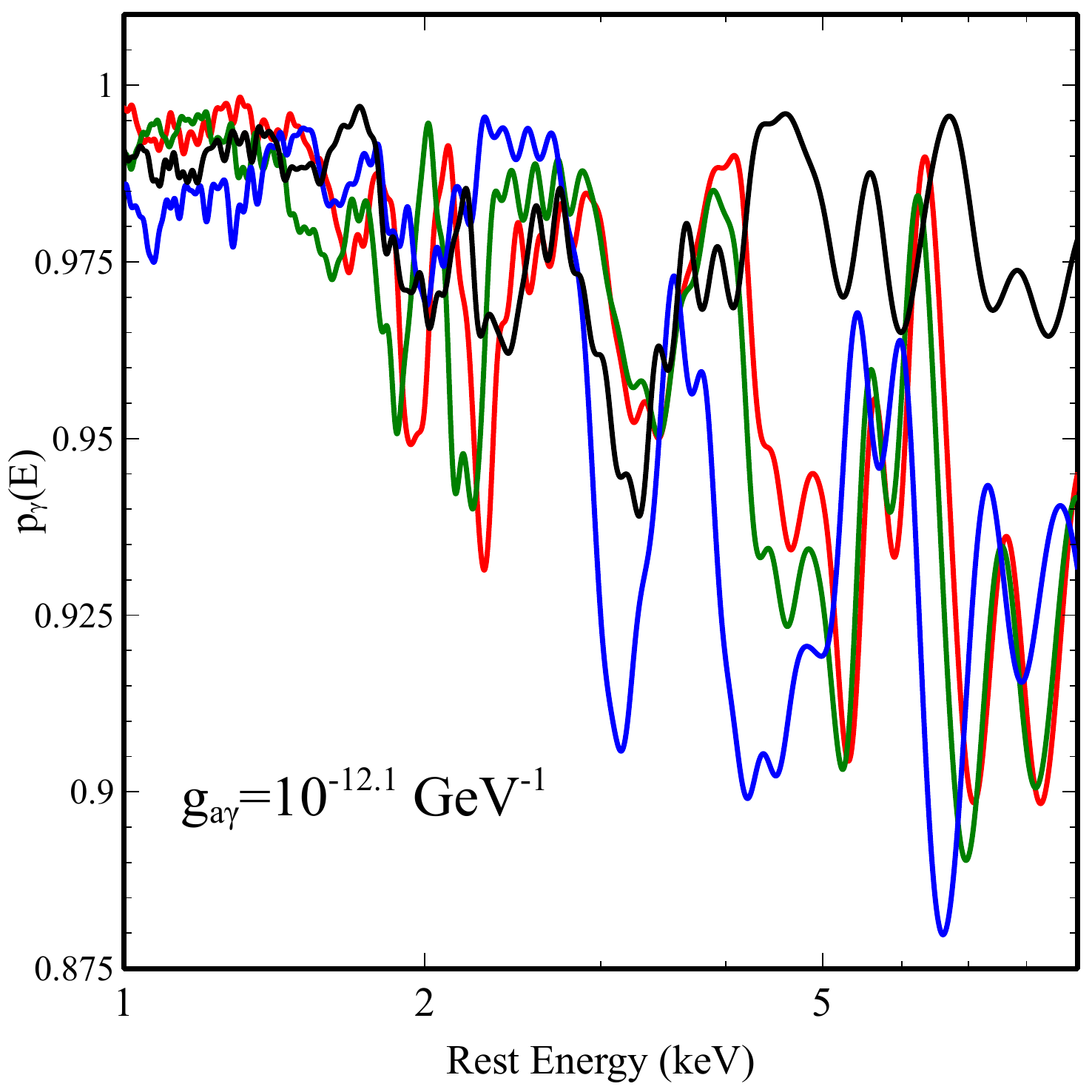}
\end{center}
\caption{Some example photon survival probability curves using one representative realization of Model-B for the magnetic field structure, $p_\gamma(E)$.  {\it Left panel : }Curves for fixed mass ($\log_{10}(m_a/\eV)=-12.7$) and magnetic field configuration, but various values of the coupling constant $\log_{10}(g_{a\gamma}/{\rm GeV}^{-1})=-11.9$ (black), $-12.1$ (blue), $-12.3$ (green), $-12.5$ (magenta), $-12.7$ (red). {\it Right panel : }Curves for fixed coupling constant $\log_{10}(g_{a\gamma}/{\rm GeV}^{-1})=-12.1$ and magnetic field configuration, but various ALP masses $\log_{10}(m_a/\eV)=-12.3$ (black), $-12.5$ (blue), $-12.7$ (green), $-12.9$ (red).}\
\label{fig:some_pe}
\vspace{0.5cm}
\end{figure*}

\noindent Our modeling of ALP spectral distortions follows that of \cite{marsh:17a}. We solve the linearized Schr\"odinger-like equation that describes the quantum mechanical oscillations between photons and ALPs as they traverse through the cluster magnetic field towards the observer. The survival probability 
of photons emitted from the nucleus that exit the cluster as photons, rather than ALPs, 
depends on the plasma density and the the cluster magnetic field. For the electron density, we use the simple analytic approximation  derived from XMM-Newton observations of the Perseus cluster  observations by  \cite{Churazov:2003hr}:
\begin{align}
n_e(r) = \frac{3.9\times10^{-2}}{\left(1+ (r/80\, {\rm kpc})^2\right)^{1.8}} + \frac{4.05\times10^{-3}}{\left(1+ (r/280\, {\rm kpc})^2\right)^{0.87}}
~{\rm cm}^{-3} \, . \label{eq:ne}
\end{align}
For the magnetic field, we use two stochastic models. \\

\noindent {\bf Model A:} We first consider a slight modification to the magnetic fields used in \cite{berg:17a}. This model is motivated by VLBA observations of the nucleus of NGC~1275 by \cite{taylor:06a}, which found Faraday rotation measures (RM) of the order of $6500$--$7500~ {\rm rad}\,  {\rm m}^{-2}$ across the tip of the southern active jet of 3C~84, the radio source associated with NGC~1275. Interpreting these RMs as arising from a narrow ($\sim 2\, $kpc) Faraday screen in the high-density ($n_e \approx 0.3\, $cm$^{-3}$) central region, \cite{taylor:06a} estimated the central magnetic field strength as $B_0 \approx25\, \mu$G. { Taking this value for the central magnetic field}, \cite{berg:17a} modeled the magnetic field along the line of sight as a series of domains.  { Within each domain, the field is taken to be randomly-oriented with a constant magnitude given by $B(r_c) = B_0 \left[n_e(r_c)/n_e(0)\right]^{0.7}$ (where $r_c$ is the radial coordinate at the center of the domain).  Motivated by the detailed RM study of Abell~2199 by \cite{vacca:12a}, the size of each domain $L$ is drawn from a random distribution with probability density proportional to $L^{-1.2}$ \citep[corresponding to the index of the 3-d RM power-spectrum found by][]{vacca:12a}, between 3.5--10\,kpc (motivated by scaling length scales in Abell~2199 to Perseus).  }


Our model A modifies this description in two ways. First, we note that the large value of $B_0$ combined with the moderate central plasma density of  eq.~\eqref{eq:ne} leads to non-negligible ALP-photon oscillations from the inner-most region of the cluster. However, the simple analytic model of  \cite{Churazov:2003hr} does not apply to small radii,  $r <10\, $kpc, where it underestimates the electron density, and where the spherically symmetric approximation is not justified. Applying eq.~\eqref{eq:ne} to this region  leads to an overestimate of the ALP-photon conversion probability. In our work, we conservatively exclude the central region, and simulate the ALP-photon oscillations from $10\, $kpc out to the virial radius, $R_{\rm vir} =1.8\, $Mpc.  Second, we note that the bulk of the ICM also acts as a Faraday screen and sources RMs 
in addition to those arising from the central region. Since this model attributes the observed RMs to the Faraday screen close to the center of the cluster, we consistently select only those magnetic field configurations in which the cluster contribution is subleading: ${\rm RM}_{\rm cluster} \leq 2000\, $rad\, m$^{-2}$. However, we have found that this restriction has no statistically significant impact on the typical conversion probabilities.

\vspace{0.5cm}

\noindent {\bf Model B:} We furthermore consider a model in which the ratio of the thermal-to-magnetic pressure is fixed to $\beta=100$ throughout the cluster.  We use the Perseus pressure profile of \cite{Fabian:2005tz} to derive a magnetic field strength of $B_{25} \approx 7.5\, \mu$G at $r=25\, $kpc. Approximating the cluster as isothermal, the magnetic field decreases with radius as $\sim \sqrt{n_e(r)}$, where we again use \eqref{eq:ne} for the electron density. { With a central field that is suppressed with respect to Model-A, the ALP-photon conversion from the central region is negligible and so we can use this model from $r=0$ to the virial radius. }
The coherence lengths of the magnetic field can be expected to grow with distance from the centre. We model this effect by drawing the coherence lengths randomly from $(1+r/50\, {\rm kpc}) \times 3.5\, $kpc to $(1+r/50\, {\rm kpc}) \times 10\, $kpc, with a power-law fall-off as $\sim L^{-1.2}$. This model produces Faraday RMs of the same order as those observed by \cite{taylor:06a}, with the cluster as the Faraday screen. { While this choice of domain-size structure is somewhat arbitrary, it is designed to allow comparison and connection with the results of Model-A as well as previous studies.}

\vspace{0.5cm}

\begin{figure}
\includegraphics[width=0.45\textwidth]{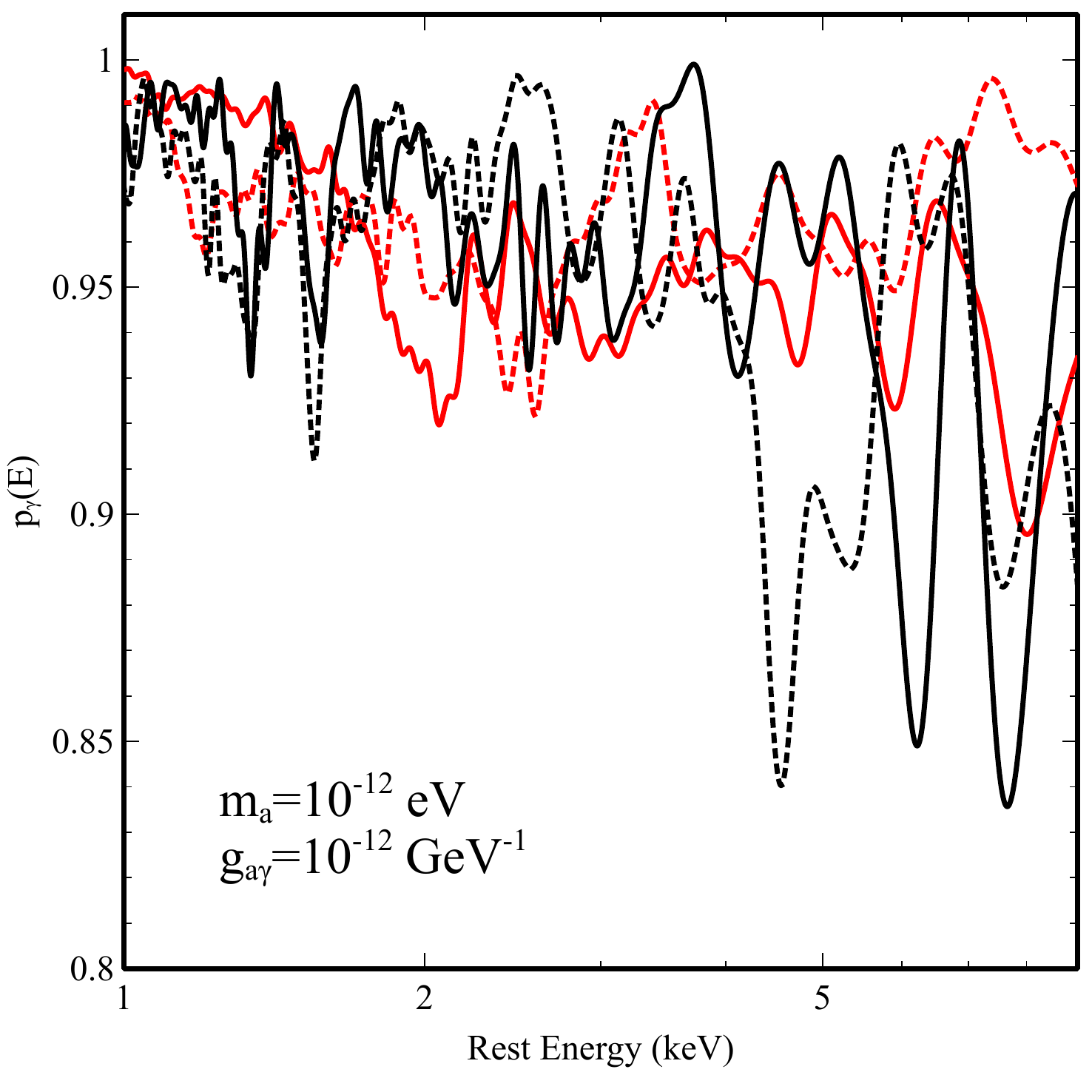}
\caption{{ Example photon survival probability curves for one choice of ALP parameters ($m_a=10^{-12}\eV, g_{a\gamma}=10^{-12}$\,GeV$^{-1}$) and two representative magnetic field realizations from each of Model-A (black) and Model-B (red). }}
\label{fig:modelab_compare}
\end{figure}

For each of our two field models, we generate 500 RM-acceptable magnetic field configurations and solve the Schr\"odinger-like equation in order to calculate photon survival probabilities across a grid of $m_a$ and $g_{a\gamma}$.  Our models sample the $(m_a, g_{a\gamma})$-plane, spanning the range $\log_{10}(m_a/\eV)\in[-13.6,-11.1]$ and $\log_{10}(g_{a\gamma}/{\rm GeV^{-1})}\in[-13,-10.7]$. The result is a library of approximately 260,000 energy-dependent photon survival probability curves for each of our two magnetic field models (Model-A and Model-B), $p_\gamma(E; m_a, g_{a\gamma}, i_{A/B})$, where $i_{A/B}$ indexes the 500 RM-acceptable magnetic field realizations for that given magnetic field model.  Some representative photon survival probability curves, and their functional dependence on $m_a$ and $g_{a\gamma}$, are shown in Figure~\ref{fig:some_pe}.  

{ In Fig.~\ref{fig:modelab_compare}, we compare photon survival probability curves from two representative realizations for each of our two magnetic field models at an illustrative point in the ALP parameter space, $m_a=10^{-12}\eV, g_{a\gamma}=10^{-12}\,{\rm GeV}^{-1}$.  Below 4\,keV, the two field models give spectral distortions of similar magnitude, although the distortions produced by Model-A are typically narrower.  Above 4\,keV, both field models show a transition to more periodic energy structures, with Model-A showing a marked increase in the magnitude of the distortions.} 

{ The domain models that we consider are simple enough to make the extensive calculations required below feasible, but complex enough to agree qualitatively with several of the features of more elaborate stochastic models in which the cluster magnetic field is taken to be a divergence-free function derived from Gaussian random fields \cite[as in][see in particular section 5.2.1]{angus:14a}. The discontinuity of the magnetic field at the boundaries of the domain does not, of course, lead to discontinuities in the conversion probability as a function of the radius. Neither the Gaussian random field model nor the discrete cell model correspond to magnetic fields that are actually realized in nature, but are simple models that capture some aspects of the underlying physical magnetic field. For photon-ALP oscillations, the relevant aspects are the (non-radial) strength of the magnetic field and its (radial) coherence length.  Testing the detailed properties of photon-to-ALP conversion in more realistic turbulent magnetic fields such as those derived from MHD simulations is an interesting exercise but beyond the scope of the current paper.} 

\section{Constraints on ALP parameters}\label{results}

\begin{figure*}
\begin{center}
\includegraphics[width=0.85\textwidth]{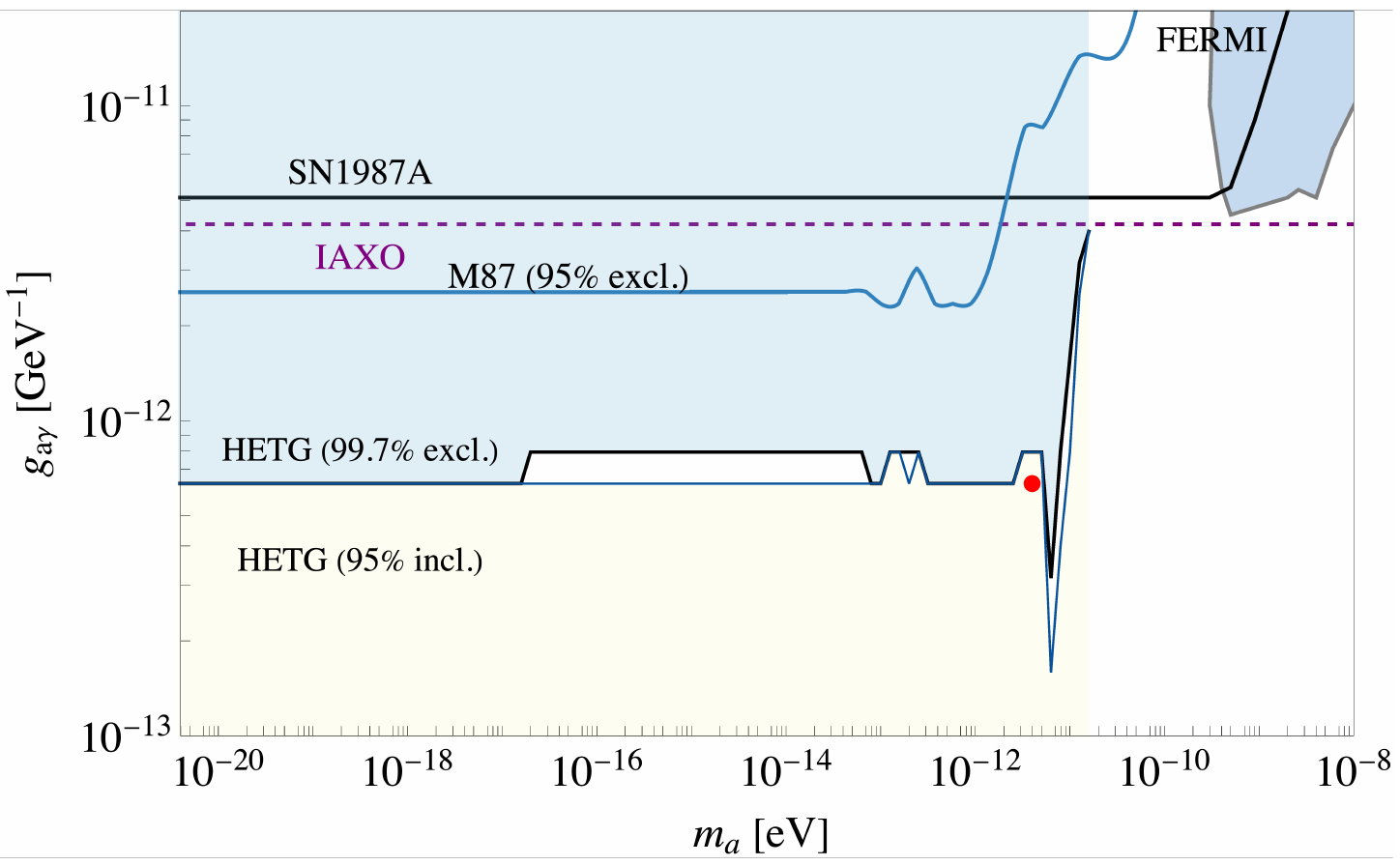}
\end{center}
\begin{center}
\includegraphics[width=0.85\textwidth]{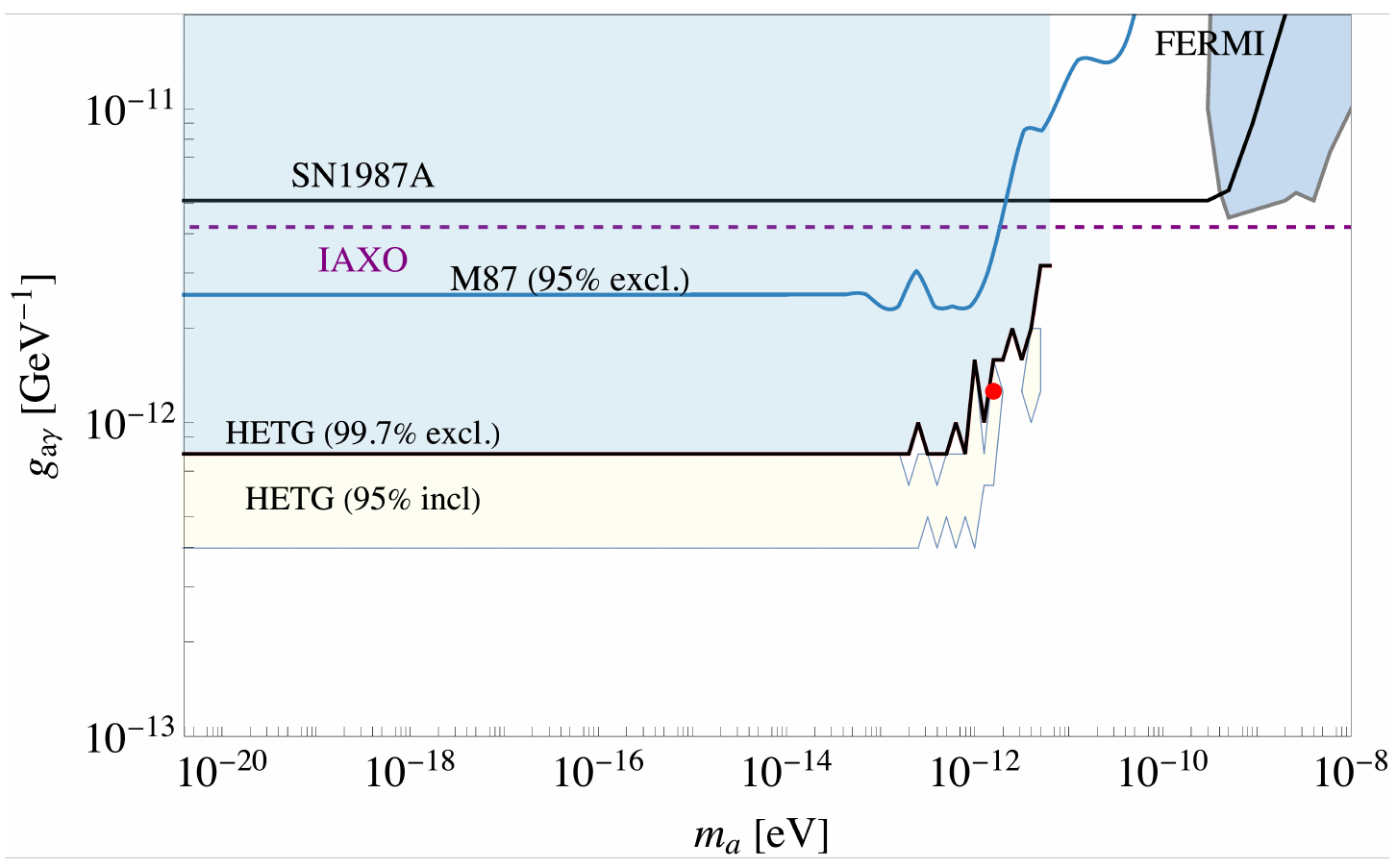}
\end{center}
\caption{Constraints on the $(m_a, g_{a\gamma})$-plane from this study for magnetic field Model-A (top) and Model-B (bottom).  We show 99.7\% confidence limit (heavy black line bounding the excluded, light blue shaded region), the 95\% confidence region (allowed region with yellow shading), and best-fit parameter values (red dots). 
Also shown are the previous most stringent 95\% limit from the \cite{marsh:17a} study of M87 (blue line), the limit from the absence of a $\gamma$-ray burst associated with SN1987 \citep{payez:15a} (black line), the limit from FERMI (gray line bounding blue excluded region at the 95\% confidence level)  as well as the projected sensitivity from the next generation helioscope IAXO.
For other, weaker limits obtained using X-ray astronomy in the same region of parameter space, see also \cite{wouters:13a, berg:17a, Conlon:2017qcw}.
}
\label{fig:results}
\vspace{0.5cm}
\end{figure*}

\noindent Equipped with the library of photon survival probability curves, we can now use our HETG spectra of NGC~1275 to determine the allowed regions of the $(m_a, g_{a\gamma})$-plane.  

In order to determine the probability of the parameters given this data, appropriately marginalised over the unknown cluster magnetic field configuration, we  follow the Bayesian procedure of \cite{marsh:17a}.  We assume flat priors on $\ln m_a$ and $\ln g_{a\gamma}$ in the range $\log_{10} (m_a/\eV)\in [-30,-11.1]$ and $\log_{10} (g_{a\gamma}/{\rm GeV}^{-1})\in [-19,-10.7]$. We will find that our results are insensitive to the minimum allowed mass, and the particular choice considered here corresponds to cosmologically large Compton wavelengths of the ALPs. The minimum allowed coupling constant corresponds to the inverse Planck mass, below which quantum gravitational corrections are expected to become important. We also assume flat priors on  the randomly generated magnetic field configurations, labelled by $i_{A/B}$.  Motivated by the initial fitting presented in Section~\ref{reduction}, our baseline spectral model for NGC~1275 consists of a power-law continuum modified by the effects of Galactic absorption \citep[$N_H=1.32\times 10^{21}\pcmsq$;][]{kalberla:05a}.  

For a given magnetic field model (Model-A and Model-B), we take each of our photon survival probability curves (indexed by $m_a$, $g_{a\gamma}$ and the magnetic field realization $i_{A/B}$), multiply by the power-law spectrum (modified by Galactic absorption), and then fit to the unbinned HEG/MEG spectra, minimizing the $C$-statistic over the HEG/MEG photon indices and HEG/MEG normalizations.  The lowest masses in our model library ($\log_{10}(m_a/\eV)=-13.6$) yield fits that are indistinguishable from the massless case, and hence these model fits are used as proxies for the very low-mass region of parameter space.  Similarly, the smallest coupling constant in our model library ($\log_{10} (g_{a\gamma}/{\rm GeV}^{-1})=-13$) are indistinguishable from the zero coupling case and hence these model fits are used as proxies for the very small coupling region of parameter space.

From the resulting set of $C(m_a, g_{a\gamma}, i_{A/B})$, we form posterior probabilities ${\cal P}_{A/B}(m_a, g_{a\gamma}, i)\propto \exp(-C/2)$ normalized such that 
\begin{equation}
\sum_{m_a, g_{a\gamma}, i_{A/B}}{\cal P}_{A/B}(m_a, g_{a\gamma}, i_{A/B})=1 \, .
\end{equation}

To obtain the posterior over the ALP-parameters alone (and account for the `look-elsewhere effect' associated with the unknown magnetic field), we then marginalize over the magnetic field configurations,
\begin{equation}
{\cal P}_{A/B}(m_a, g_{a\gamma})=\sum_{i_{A/B}}{\cal P}_{A/B}(m_a, g_{a\gamma}, i_{A/B}).
\end{equation}
{ The maximum value of ${\cal P}_{A/B}(m_a, g_{a\gamma})$  gives the best-fit values of the ALP parameters.  Contours of equal ${\cal P}_{A/B}(m_a, g_{a\gamma})$ provide the boundaries of the credible regions of the ALP parameters. The $x$\% credible region includes the points with largest  ${\cal P}_{A/B}(m_a, g_{a\gamma})$ so that their sum accounts for $x$\% of the total probability.  For the marginalized probabilities of this work, this method presents no ambiguities.}

Figure~\ref{fig:results} shows the main result of this paper, our new constraints on the $(m_a, g_{a\gamma})$-plane.  { For reference, Fig.~\ref{fig:results} also shows the previously tightest constraints on ALPs in this mass range from SN1987A \citep{payez:15a}, and the lack of spectral distortions in the X-ray spectrum of M87 \citep{marsh:17a} and the {\it Fermi}/$\gamma$-ray spectrum of NGC~1275 \citep{ajello:16a}.}  The posterior {probability from our new analysis} over the ALP parameters has a complex structure.  For $m_a<1\times 10^{-12}\eV$, we set a strong upper limit of $g_{a\gamma}<6.3\times 10^{-13}\,{\rm GeV}^{-1}$ (model-A) and $g_{a\gamma}<7.9\times 10^{-13}\,{\rm GeV}^{-1}$ (model-B) at the 99.7\%  level.  This stems directly from the fact that the HEG/MEG spectra display no significant spectral distortions exceeding $\pm 3\%$.  

At the 95\% level and for model-B, the marginalization process picks out a preferred value of $g_{a\gamma}\approx 4-8\times 10^{-13}\,{\rm GeV}^{-1}$.\footnote{We note that the strength of the preference for a non-vanishing value of the coupling $g_{a\gamma}$ is prior-dependent: should we have restricted the prior range to only the sampled region of $\log_{10}(g_{a\gamma}/{\rm GeV^{-1})}\in[-13,-10.7]$, the 99.7\% confidence regions obtained from either of the  magnetic field models would exclude a vanishing ALP-photon coupling. In contrast, the upper limit on $g_{a\gamma}$ is  highly prior-independent, and  the large prior range considered in this paper is therefore conservative.}  An examination of a sample of the photon survival probability curves for the most probable models shows that they all share a modest dip at 1.2\,keV and a broad hump at 3--4\,keV, structure that is visually apparent in the spectra (Fig.~\ref{fig:plratio}). At these levels, however, such spectral structures could easily result from remaining calibration errors in the HETG energy-dependence effective area \citep{marshall:12a} and thus we cannot claim even a tentative detection of ALPs on the basis of the enclosed 95\% confidence contour.  

{We end this section with a brief discussion of the Bayesian evidence for ALP models compared with a no-ALP hypothesis.  For magnetic field Model-A, we obtain a Bayes factor of $K=1.5$, ``barely worth mentioning'' on the Jeffrey's scale.  For Model-B, we obtain $K=22.8$ which constitutes ``strong evidence'' for ALPs on the Jeffrey's scale. This is a restatement of the fact that the data does, indeed, possess structure that can be fitted by ALP distortions. }

\section{Discussion and Conclusions}\label{conclusions}

\noindent A deep exposure of NGC~1275 at the center of the Perseus cluster of galaxies with the {\it Chandra}/HETG has allowed us to obtain the highest quality spectrum of this AGN free from strong contamination by the surrounding cluster emission and free from the effects of photon pileup.  Apart from subtle structure in the iron-band (6--7\,keV), we find that the 1--9\,keV spectrum of NGC~1275 is accurately described by a power-law continuum form modified by the effects of modest Galactic absorption; deviations from the power-law are at the $\pm 3\%$ level.  Taking this to be the intrinsic spectrum of the AGN, we proceed to use these data to constrain models for ALP-photon oscillations in the magnetic field of the Perseus cluster.  We have obtained the most stringent limit yet on the ALP-photon coupling constant of very light ALPs with masses $m_a<1\times 10^{-12}\eV$:
\begin{equation}
g_{a\gamma} <7.9\times 10^{-13}\,{\rm GeV}^{-1},\hspace{0.5cm} \hbox{(99.7\% confidence)},
\end{equation}
for magnetic field model-B, with an even more stringent limit of $g_{a\gamma}<6.3\times 10^{-13}\,{\rm GeV}^{-1}$ for model-A.  Even at this much higher level of confidence, our limits are $3-4\times$ stronger than those obtained by \cite{marsh:17a} for M87/Virgo, and over $5\times $ stronger than the those found by \cite{berg:17a} with non-dispersive spectroscopy of NGC~1275/Perseus.

For one of our two magnetic field models, the ALP conversion models pick out some remaining structure in the HEG/MEG spectra and hence, at the 95\% level (but not the 99.7\% level) lock onto a preferred non-zero coupling constant of $g_{a\gamma}\approx 4-8\times 10^{-13}\,{\rm GeV}^{-1}$.  However, we acknowledge that $\pm 3\%$ is very plausibly at the level of residual calibration uncertainties in the HETG energy-dependent effective area and hence this result must be viewed with extreme caution.  

For these very light ALPs ($m_a<1\times 10^{-11}\eV$) our astrophysical limits are now tighter than those obtained by the fact that SN1987A did not generate a gamma-ray burst via the ALP-mediated escape of gamma-rays from the collapsing stellar core \citep{payez:15a}.  Our limits have also exceeded the projected sensitivity of the next-generation helioscope, the International Axion Observatory (IAXO) \citep{armengaud:14a}, as well as the next generation light-shining-through-walls experiments such as ALPS-II \citep{bahre:13a}.

While the focus of this paper are the constraints on ALPs, we note that our data allow a strong test of the \cite{conlon:17a} fluorescent dark matter model. These authors note that there are hints of an absorption feature at 3.5\,keV in the {\it XMM-Newton}/EPIC-pn of NGC~1275. In an attempt to reconcile the claim of a 3.5\,keV dark matter emission line in the {\it XMM-Newton} spectrum of the Perseus ICM \citep{bulbul:14a,boyarsky:14a} with the non-detection of any such feature in the {\it Hitomi} spectrum of the the system \citep{hitomi:17a}, \cite{conlon:17a} proceed to formulate a two-level dark matter model in which a dark matter absorption line in the AGN spectrum offsets the dark matter emission line from the cluster in system-integrated spectra such as that produced by {\it Hitomi}.  This hypothesis requires that the dark matter imprints an absorption line close to 3.5\,keV with an equivalent width of 15\, eV.   Our HETG current spectrum sets an upper limit of $4\eV$ with 99\% confidence, allowing us to rule out this version of the fluorescent dark matter model.

Following on from \cite{wouters:13a}, \cite{berg:17a} and \cite{marsh:17a}, our work is just the latest to highlight the power of X-ray transparency studies of galaxy clusters to probe the physics of ALPs.  Equipped with high-resolution, non-dispersion, high count rate micro-calorimater arrays at the focus of X-ray telescopes with $A>1\m^2$, the next generation of X-ray flagship missions {\it Athena} and {\it Lynx} will dramatically advance the quality of the spectrum possible for an embedded AGN such as NGC~1275.  

But the current study also points to the challenges that need to be overcome if we are extend this technique further and dig deeper into the parameter space. Even if photon pileup/deadtime and contamination by ICM emission is rendered negligible, the relative calibration of the energy-dependent effective area of the X-ray spectrometer will set a floor on the sensitivity of any studies based on spectral distortions irrespective of the photon statistics.  For example, for a representative Perseus magnetic field model, a very light ALP with $g_{a\gamma} =3\times 10^{-13}\,{\rm GeV}^{-1}$ produces distortions at the 1.5\% (3\%) level in the $E\sim 2-4\keV (5-8\keV)$ band.   For $g_{a\gamma} =2\times 10^{-13}\,{\rm GeV}^{-1}$, the corresponding distortion is 0.75\% (1.5\%).  Thus, to significantly further this technique for searching for ALPs, future X-ray spectrometers must achieve relative effective area calibrations of 1\% or better. It is clear that in order to push the ALP constraints significantly further than we have done here (or actually obtain a robust detection of very light ALPs) will require advances in broad-band, on-orbit, relative calibration of future X-ray spectrometers.

{Should future studies actually detect spectral distortions due to ALPs, the recovered ALP parameters $(m_a, g_{a\gamma})$ will clearly depend upon the magnetic field model assumed in the analysis.  As an illustration of this issue, we have created simulated 490\,ks HEG/MEG spectra of NGC~1275 modified by ALPs with $\log(m_a/{\rm eV})=-12.5$ and $\log(g_{a\gamma}/{\rm GeV}^{-1})=-12$ employing one realization of magnetic field Model-A. We then ran the simulated data through our full analysis pipeline using magnetic field Model-B.  The ALP signal is detected at the 99.7\% level, with $g_{a\gamma}$ constrained to 0.2\,dex albeit biased low compared with the true value by $\sim$0.1\,dex.  The 68\% constraints on the mass are $\log(m_a/{\rm eV})\in[-12.3,-12.1]$, slightly higher than the injected signal, but the mass is unconstrained at the 99.7\% level.  This simple exercise highlights the importance of a high-quality magnetic field model, and hence the need for more detailed RM mapping of target clusters.} 

\section*{Acknowledgements}

We thank the Chandra Science Center, and especially Hermann Marshall, for advice and guidance in the execution of these observations. D.M.~thanks Jens Jache for discussions. We also thank the anonymous referee for their suggestions and insights that improved the manuscript.  C.S.R. thanks the UK Science and Technology Facilities Council (STFC) for support under the New Applicant grant ST/R000867/1, and the European Research Council (ERC) for support under the European Union's Horizon 2020 research and innovation programme (grant 834203).  D.M. acknowledges support from the Swedish Science Agency grant 2018-03641 and from European Research Council Grant 742104. H.R.R acknowledges support from an STFC Rutherford Fellowship and an Anne McLaren Fellowship.  A.C.F. acknowledges support from ERC Advanced Grant 340442.  F.T. acknowledges support by the Programma per Giovani Ricercatori - anno 2014 `Rita Levi Montalcini'.  S.V. acknowledges support from a Raymond and Beverley Sackler Distinguished Visitor Fellowship and thanks the host Institute, the Institute of Astronomy, where this work was concluded. S.V. also acknowledges support by the Science and Technology Facilities Council (STFC) and by the Kavli Institute for Cosmology, Cambridge.  S.V. also thanks support from NASA under the Chandra Guest Observer program.  This work used, in part, the Cambridge COSMOS SMP system, part of the STFC DiRAC HPC Facility supported by BIS NeI capital grant ST/J005673/1 and STFC grants ST/H008586/1, ST/K00333X/1.


\begin{thebibliography}{}
\expandafter\ifx\csname natexlab\endcsname\relax\def\natexlab#1{#1}\fi
\providecommand{\url}[1]{\href{#1}{#1}}
\providecommand{\dodoi}[1]{doi:~\href{http://doi.org/#1}{\nolinkurl{#1}}}
\providecommand{\doeprint}[1]{\href{http://ascl.net/#1}{\nolinkurl{http://ascl.net/#1}}}
\providecommand{\doarXiv}[1]{\href{https://arxiv.org/abs/#1}{\nolinkurl{https://arxiv.org/abs/#1}}}

\bibitem[{{Abbott} \& {Sikivie}(1983)}]{abbott:83a}
{Abbott}, L.~F., \& {Sikivie}, P. 1983, Physics Letters B, 120, 133,
  \dodoi{10.1016/0370-2693(83)90638-X}

\bibitem[{{Ajello} {et~al.}(2016){Ajello}, {Albert}, {Anderson}, {Baldini},
  {Barbiellini}, {Bastieri}, {Bellazzini}, {Bissaldi}, {Blandford}, {Bloom},
  {Bonino}, {Bottacini}, {Bregeon}, {Bruel}, {Buehler}, {Caliandro}, {Cameron},
  {Caragiulo}, {Caraveo}, {Cecchi}, {Chekhtman}, {Ciprini}, {Cohen-Tanugi},
  {Conrad}, {Costanza}, {D'Ammando}, {de Angelis}, {de Palma}, {Desiante}, {Di
  Mauro}, {Di Venere}, {Dom{\'\i}nguez}, {Drell}, {Favuzzi}, {Focke},
  {Franckowiak}, {Fukazawa}, {Funk}, {Fusco}, {Gargano}, {Gasparrini},
  {Giglietto}, {Glanzman}, {Godfrey}, {Guiriec}, {Horan}, {J{\'o}hannesson},
  {Katsuragawa}, {Kensei}, {Kuss}, {Larsson}, {Latronico}, {Li}, {Li}, {Longo},
  {Loparco}, {Lubrano}, {Madejski}, {Maldera}, {Manfreda}, {Mayer},
  {Mazziotta}, {Meyer}, {Michelson}, {Mirabal}, {Mizuno}, {Monzani},
  {Morselli}, {Moskalenko}, {Murgia}, {Negro}, {Nuss}, {Okada}, {Orlando},
  {Ormes}, {Paneque}, {Perkins}, {Pesce-Rollins}, {Piron}, {Pivato}, {Porter},
  {Rain{\`o}}, {Rando}, {Razzano}, {Reimer}, {S{\'a}nchez-Conde}, {Sgr{\`o}},
  {Simone}, {Siskind}, {Spada}, {Spandre}, {Spinelli}, {Takahashi}, {Thayer},
  {Torres}, {Tosti}, {Troja}, {Uchiyama}, {Wood}, {Wood}, {Zaharijas},
  {Zimmer}, \& {Fermi-LAT Collaboration}}]{ajello:16a}
{Ajello}, M., {Albert}, A., {Anderson}, B., {et~al.} 2016, \prl, 116, 161101,
  \dodoi{10.1103/PhysRevLett.116.161101}

\bibitem[{{Angus} {et~al.}(2014){Angus}, {Conlon}, {Marsh}, {Powell}, \&
  {Witkowski}}]{angus:14a}
{Angus}, S., {Conlon}, J.~P., {Marsh}, M.~C.~D., {Powell}, A.~J., \&
  {Witkowski}, L.~T. 2014, \jcap, 2014, 026,
  \dodoi{10.1088/1475-7516/2014/09/026}

\bibitem[{{Armengaud} {et~al.}(2014){Armengaud}, {Avignone}, {Betz}, {Brax},
  {Brun}, {Cantatore}, {Carmona}, {Carosi}, {Caspers}, {Caspi}, {Cetin},
  {Chelouche}, {Christensen}, {Dael}, {Dafni}, {Davenport}, {Derbin}, {Desch},
  {Diago}, {D{\"o}brich}, {Dratchnev}, {Dudarev}, {Eleftheriadis},
  {Fanourakis}, {Ferrer-Ribas}, {Gal{\'a}n}, {Garc{\'{\i}}a}, {Garza},
  {Geralis}, {Gimeno}, {Giomataris}, {Gninenko}, {G{\'o}mez},
  {Gonz{\'a}lez-D{\'{\i}}az}, {Guendelman}, {Hailey}, {Hiramatsu}, {Hoffmann},
  {Horns}, {Iguaz}, {Irastorza}, {Isern}, {Imai}, {Jakobsen}, {Jaeckel},
  {Jakov{\v c}i{\'c}}, {Kaminski}, {Kawasaki}, {Karuza}, {Kr{\v c}mar},
  {Kousouris}, {Krieger}, {Laki{\'c}}, {Limousin}, {Lindner}, {Liolios},
  {Luz{\'o}n}, {Matsuki}, {Muratova}, {Nones}, {Ortega}, {Papaevangelou},
  {Pivovaroff}, {Raffelt}, {Redondo}, {Ringwald}, {Russenschuck}, {Ruz},
  {Saikawa}, {Savvidis}, {Sekiguchi}, {Semertzidis}, {Shilon}, {Sikivie},
  {Silva}, {ten Kate}, {Tomas}, {Troitsky}, {Vafeiadis}, {van Bibber},
  {Vedrine}, {Villar}, {Vogel}, {Walckiers}, {Weltman}, {Wester}, {Yildiz}, \&
  {Zioutas}}]{armengaud:14a}
{Armengaud}, E., {Avignone}, F.~T., {Betz}, M., {et~al.} 2014, Journal of
  Instrumentation, 9, T05002, \dodoi{10.1088/1748-0221/9/05/T05002}

\bibitem[{{Arnaud}(1996)}]{arnaud:96a}
{Arnaud}, K.~A. 1996, in Astronomical Society of the Pacific Conference Series,
  Vol. 101, Astronomical Data Analysis Software and Systems V, ed. G.~H.
  {Jacoby} \& J.~{Barnes}, 17

\bibitem[{{B{\"a}hre} {et~al.}(2013){B{\"a}hre}, {D{\"o}brich},
  {Dreyling-Eschweiler}, {Ghazaryan}, {Hodajerdi}, {Horns}, {Januschek},
  {Knabbe}, {Lindner}, {Notz}, {Ringwald}, {von Seggern}, {Stromhagen},
  {Trines}, \& {Willke}}]{bahre:13a}
{B{\"a}hre}, R., {D{\"o}brich}, B., {Dreyling-Eschweiler}, J., {et~al.} 2013,
  Journal of Instrumentation, 8, T09001, \dodoi{10.1088/1748-0221/8/09/T09001}

\bibitem[{{Berg} {et~al.}(2017){Berg}, {Conlon}, {Day}, {Jennings},
  {Krippendorf}, {Powell}, \& {Rummel}}]{berg:17a}
{Berg}, M., {Conlon}, J.~P., {Day}, F., {et~al.} 2017, \apj, 847, 101,
  \dodoi{10.3847/1538-4357/aa8b16}

\bibitem[{{Boyarsky} {et~al.}(2014){Boyarsky}, {Ruchayskiy}, {Iakubovskyi}, \&
  {Franse}}]{boyarsky:14a}
{Boyarsky}, A., {Ruchayskiy}, O., {Iakubovskyi}, D., \& {Franse}, J. 2014,
  Physical Review Letters, 113, 251301, \dodoi{10.1103/PhysRevLett.113.251301}

\bibitem[{{Brockway} {et~al.}(1996){Brockway}, {Carlson}, \&
  {Raffelt}}]{brockway:96a}
{Brockway}, J.~W., {Carlson}, E.~D., \& {Raffelt}, G.~G. 1996, Physics Letters
  B, 383, 439, \dodoi{10.1016/0370-2693(96)00778-2}

\bibitem[{{Bulbul} {et~al.}(2014){Bulbul}, {Markevitch}, {Foster}, {Smith},
  {Loewenstein}, \& {Randall}}]{bulbul:14a}
{Bulbul}, E., {Markevitch}, M., {Foster}, A., {et~al.} 2014, \apj, 789, 13,
  \dodoi{10.1088/0004-637X/789/1/13}

\bibitem[{{Caputo} {et~al.}(2019){Caputo}, {Sberna}, {Fr{\'\i}as}, {Blas},
  {Pani}, {Shao}, \& {Yan}}]{caputo:19a}
{Caputo}, A., {Sberna}, L., {Fr{\'\i}as}, M., {et~al.} 2019, \prd, 100, 063515,
  \dodoi{10.1103/PhysRevD.100.063515}

\bibitem[{Chen \& Conlon(2018)}]{Chen:2017mjf}
Chen, L., \& Conlon, J.~P. 2018, Mon. Not. Roy. Astron. Soc., 479, 2243,
  \dodoi{10.1093/mnras/sty1591}

\bibitem[{Churazov {et~al.}(2003)Churazov, Forman, Jones, \&
  Bohringer}]{Churazov:2003hr}
Churazov, E., Forman, W., Jones, C., \& Bohringer, H. 2003, Astrophys. J., 590,
  225, \dodoi{10.1086/374923}

\bibitem[{Conlon {et~al.}(2017)Conlon, Day, Jennings, Krippendorf, \&
  Rummel}]{Conlon:2017qcw}
Conlon, J.~P., Day, F., Jennings, N., Krippendorf, S., \& Rummel, M. 2017,
  JCAP, 1707, 005, \dodoi{10.1088/1475-7516/2017/07/005}

\bibitem[{{Conlon} {et~al.}(2017){Conlon}, {Day}, {Jennings}, {Krippendorf}, \&
  {Rummel}}]{conlon:17a}
{Conlon}, J.~P., {Day}, F., {Jennings}, N., {Krippendorf}, S., \& {Rummel}, M.
  2017, \prd, 96, 123009, \dodoi{10.1103/PhysRevD.96.123009}

\bibitem[{{Davis}(2001)}]{davis:01a}
{Davis}, J.~E. 2001, \apj, 562, 575, \dodoi{10.1086/323488}

\bibitem[{{Dine} \& {Fischler}(1983)}]{dine:83a}
{Dine}, M., \& {Fischler}, W. 1983, Physics Letters B, 120, 137,
  \dodoi{10.1016/0370-2693(83)90639-1}

\bibitem[{Fabian {et~al.}(2006)Fabian, Sanders, Taylor, Allen, Crawford,
  Johnstone, \& Iwasawa}]{Fabian:2005tz}
Fabian, A.~C., Sanders, J.~S., Taylor, G.~B., {et~al.} 2006, Mon. Not. Roy.
  Astron. Soc., 366, 417, \dodoi{10.1111/j.1365-2966.2005.09896.x}

\bibitem[{{Fabian} {et~al.}(2000){Fabian}, {Sanders}, {Ettori}, {Taylor},
  {Allen}, {Crawford}, {Iwasawa}, {Johnstone}, \& {Ogle}}]{fabian:00a}
{Fabian}, A.~C., {Sanders}, J.~S., {Ettori}, S., {et~al.} 2000, \mnras, 318,
  L65, \dodoi{10.1046/j.1365-8711.2000.03904.x}

\bibitem[{{Graham} {et~al.}(2015){Graham}, {Irastorza}, {Lamoreaux}, {Lindner},
  \& {van Bibber}}]{graham:15a}
{Graham}, P.~W., {Irastorza}, I.~G., {Lamoreaux}, S.~K., {Lindner}, A., \& {van
  Bibber}, K.~A. 2015, Annual Review of Nuclear and Particle Science, 65, 485

\bibitem[{{Grifols} {et~al.}(1996){Grifols}, {Mass{\'o}}, \&
  {Toldr{\`a}}}]{grifols:96a}
{Grifols}, J.~A., {Mass{\'o}}, E., \& {Toldr{\`a}}, R. 1996, \prl, 77, 2372,
  \dodoi{10.1103/PhysRevLett.77.2372}

\bibitem[{{Hitomi Collaboration} {et~al.}(2017){Hitomi Collaboration},
  {Aharonian}, {Akamatsu}, {Akimoto}, {Allen}, {Angelini}, {Arnaud}, {Audard},
  {Awaki}, {Axelsson}, {Bamba}, \& et~al.}]{hitomi:17a}
{Hitomi Collaboration}, {Aharonian}, F.~A., {Akamatsu}, H., {et~al.} 2017,
  \apjl, 837, L15, \dodoi{10.3847/2041-8213/aa61fa}

\bibitem[{{Hitomi Collaboration} {et~al.}(2018){Hitomi Collaboration},
  {Aharonian}, {Akamatsu}, {Akimoto}, {Allen}, {Angelini}, {Audard}, {Awaki},
  {Axelsson}, {Bamba}, {Bautz}, {Blandford}, {Brenneman}, {Brown}, {Bulbul},
  {Cackett}, {Chernyakova}, {Chiao}, {Coppi}, {Costantini}, {de Plaa}, {de
  Vries}, {den Herder}, {Done}, {Dotani}, {Ebisawa}, {Eckart}, {Enoto}, {Ezoe},
  {Fabian}, {Ferrigno}, {Foster}, {Fujimoto}, {Fukazawa}, {Furuzawa},
  {Galeazzi}, {Gallo}, {Gandhi}, {Giustini}, {Goldwurm}, {Gu}, {Guainazzi},
  {Haba}, {Hagino}, {Hamaguchi}, {Harrus}, {Hatsukade}, {Hayashi}, {Hayashi},
  {Hayashida}, {Hiraga}, {Hornschemeier}, {Hoshino}, {Hughes}, {Ichinohe},
  {Iizuka}, {Inoue}, {Inoue}, {Ishida}, {Ishikawa}, {Ishisaki}, {Iwai},
  {Kaastra}, {Kallman}, {Kamae}, {Kataoka}, {Katsuda}, {Kawai}, {Kelley},
  {Kilbourne}, {Kitaguchi}, {Kitamoto}, {Kitayama}, {Kohmura}, {Kokubun},
  {Koyama}, {Koyama}, {Kretschmar}, {Krimm}, {Kubota}, {Kunieda}, {Laurent},
  {Lee}, {Leutenegger}, {Limousin}, {Loewenstein}, {Long}, {Lumb}, {Madejski},
  {Maeda}, {Maier}, {Makishima}, {Markevitch}, {Matsumoto}, {Matsushita},
  {McCammon}, {McNamara}, {Mehdipour}, {Miller}, {Miller}, {Mineshige},
  {Mitsuda}, {Mitsuishi}, {Miyazawa}, {Mizuno}, {Mori}, {Mori}, {Mukai},
  {Murakami}, {Mushotzky}, {Nakagawa}, {Nakajima}, {Nakamori}, {Nakashima},
  {Nakazawa}, {Nobukawa}, {Nobukawa}, {Noda}, {Odaka}, {Ohashi}, {Ohno},
  {Okajima}, {Ota}, {Ozaki}, {Paerels}, {Paltani}, {Petre}, {Pinto}, {Porter},
  {Pottschmidt}, {Reynolds}, {Safi-Harb}, {Saito}, {Sakai}, {Sasaki}, {Sato},
  {Sato}, {Sato}, {Sawada}, {Schartel}, {Serlemitsos}, {Seta}, {Shidatsu},
  {Simionescu}, {Smith}, {Soong}, {Stawarz}, {Sugawara}, {Sugita},
  {Szymkowiak}, {Tajima}, {Takahashi}, {Takahashi}, {Takeda}, {Takei},
  {Tamagawa}, {Tamura}, {Tanaka}, {Tanaka}, {Tanaka}, {Tashiro}, {Tawara},
  {Terada}, {Terashima}, {Tombesi}, {Tomida}, {Tsuboi}, {Tsujimoto}, {Tsunemi},
  {Tsuru}, {Uchida}, {Uchiyama}, {Uchiyama}, {Ueda}, {Ueda}, {Uno}, {Urry},
  {Ursino}, {Watanabe}, {Werner}, {Wilkins}, {Williams}, {Yamada}, {Yamaguchi},
  {Yamaoka}, {Yamasaki}, {Yamauchi}, {Yamauchi}, {Yaqoob}, {Yatsu}, {Yonetoku},
  {Zhuravleva}, {Zoghbi}, \& {Kawamuro}}]{hitomi:18a}
{Hitomi Collaboration}, {Aharonian}, F., {Akamatsu}, H., {et~al.} 2018, \pasj,
  70, 13, \dodoi{10.1093/pasj/psx147}

\bibitem[{{Irastorza} \& {Redondo}(2018)}]{irastorza:18a}
{Irastorza}, I.~G., \& {Redondo}, J. 2018, Progress in Particle and Nuclear
  Physics, 102, 89

\bibitem[{{Kalberla} {et~al.}(2005){Kalberla}, {Burton}, {Hartmann}, {Arnal},
  {Bajaja}, {Morras}, \& {P{\"o}ppel}}]{kalberla:05a}
{Kalberla}, P.~M.~W., {Burton}, W.~B., {Hartmann}, D., {et~al.} 2005, \aap,
  440, 775, \dodoi{10.1051/0004-6361:20041864}

\bibitem[{{Kohri} \& {Kodama}(2017)}]{kohri:17a}
{Kohri}, K., \& {Kodama}, H. 2017, \prd, 96, 051701,
  \dodoi{10.1103/PhysRevD.96.051701}

\bibitem[{{Marsh} {et~al.}(2017){Marsh}, {Russell}, {Fabian}, {McNamara},
  {Nulsen}, \& {Reynolds}}]{marsh:17a}
{Marsh}, M.~C.~D., {Russell}, H.~R., {Fabian}, A.~C., {et~al.} 2017, JCAP, 12,
  036, \dodoi{10.1088/1475-7516/2017/12/036}

\bibitem[{{Marshall}(2012)}]{marshall:12a}
{Marshall}, H.~L. 2012, in Society of Photo-Optical Instrumentation Engineers
  (SPIE) Conference Series, Vol. 8443, \procspie, 844348,
  \dodoi{10.1117/12.927209}

\bibitem[{{Mukherjee} {et~al.}(2019){Mukherjee}, {Spergel}, {Khatri}, \& {Wand
  elt}}]{mukherjee:19a}
{Mukherjee}, S., {Spergel}, D.~N., {Khatri}, R., \& {Wand elt}, B.~D. 2019,
  arXiv e-prints, arXiv:1908.07534.
\newblock \doarXiv{1908.07534}

\bibitem[{{Payez} {et~al.}(2015){Payez}, {Evoli}, {Fischer}, {Giannotti},
  {Mirizzi}, \& {Ringwald}}]{payez:15a}
{Payez}, A., {Evoli}, C., {Fischer}, T., {et~al.} 2015, JCAP, 2, 006,
  \dodoi{10.1088/1475-7516/2015/02/006}

\bibitem[{{Peccei} \& {Quinn}(1977)}]{peccei:77a}
{Peccei}, R.~D., \& {Quinn}, H.~R. 1977, Physical Review Letters, 38, 1440,
  \dodoi{10.1103/PhysRevLett.38.1440}

\bibitem[{{Planck Collaboration} {et~al.}(2014){Planck Collaboration}, {Ade},
  {Aghanim}, {Armitage-Caplan}, {Arnaud}, {Ashdown}, {Atrio-Barandela},
  {Aumont}, {Baccigalupi}, {Banday}, \& et~al.}]{ade:14a}
{Planck Collaboration}, {Ade}, P.~A.~R., {Aghanim}, N., {et~al.} 2014, \aap,
  571, A16, \dodoi{10.1051/0004-6361/201321591}

\bibitem[{Preskill {et~al.}(1983)Preskill, Wise, \& Wilczek}]{Preskill:1982cy}
Preskill, J., Wise, M.~B., \& Wilczek, F. 1983, Phys. Lett., B120, 127,
  \dodoi{10.1016/0370-2693(83)90637-8}

\bibitem[{{Tanabashi} {et~al.}(2018){Tanabashi}, {Hagiwara}, {Hikasa},
  {Nakamura}, {Sumino}, {Takahashi}, {Tanaka}, {Agashe}, {Aielli}, {Amsler},
  {Antonelli}, {Asner}, {Baer}, {Banerjee}, {Barnett}, {Basaglia}, {Bauer},
  {Beatty}, {Belousov}, {Beringer}, {Bethke}, {Bettini}, {Bichsel}, {Biebel},
  {Black}, {Blucher}, {Buchmuller}, {Burkert}, {Bychkov}, {Cahn}, {Carena},
  {Ceccucci}, {Cerri}, {Chakraborty}, {Chen}, {Chivukula}, {Cowan}, {Dahl},
  {D'Ambrosio}, {Damour}, {de Florian}, {de Gouv{\^e}a}, {DeGrand}, {de Jong},
  {Dissertori}, {Dobrescu}, {D'Onofrio}, {Doser}, {Drees}, {Dreiner}, {Dwyer},
  {Eerola}, {Eidelman}, {Ellis}, {Erler}, {Ezhela}, {Fetscher}, {Fields},
  {Firestone}, {Foster}, {Freitas}, {Gallagher}, {Garren}, {Gerber}, {Gerbier},
  {Gershon}, {Gershtein}, {Gherghetta}, {Godizov}, {Goodman}, {Grab},
  {Gritsan}, {Grojean}, {Groom}, {Gr{\"u}newald}, {Gurtu}, {Gutsche}, {Haber},
  {Hanhart}, {Hashimoto}, {Hayato}, {Hayes}, {Hebecker}, {Heinemeyer},
  {Heltsley}, {Hern{\'a}ndez-Rey}, {Hisano}, {H{\"o}cker}, {Holder},
  {Holtkamp}, {Hyodo}, {Irwin}, {Johnson}, {Kado}, {Karliner}, {Katz}, {Klein},
  {Klempt}, {Kowalewski}, {Krauss}, {Kreps}, {Krusche}, {Kuyanov}, {Kwon},
  {Lahav}, {Laiho}, {Lesgourgues}, {Liddle}, {Ligeti}, {Lin}, {Lippmann},
  {Liss}, {Littenberg}, {Lugovsky}, {Lugovsky}, {Lusiani}, {Makida}, {Maltoni},
  {Mannel}, {Manohar}, {Marciano}, {Martin}, {Masoni}, {Matthews},
  {Mei{\ss}ner}, {Milstead}, {Mitchell}, {M{\"o}nig}, {Molaro}, {Moortgat},
  {Moskovic}, {Murayama}, {Narain}, {Nason}, {Navas}, {Neubert}, {Nevski},
  {Nir}, {Olive}, {Pagan Griso}, {Parsons}, {Patrignani}, {Peacock},
  {Pennington}, {Petcov}, {Petrov}, {Pianori}, {Piepke}, {Pomarol}, {Quadt},
  {Rademacker}, {Raffelt}, {Ratcliff}, {Richardson}, {Ringwald}, {Roesler},
  {Rolli}, {Romaniouk}, {Rosenberg}, {Rosner}, {Rybka}, {Ryutin}, {Sachrajda},
  {Sakai}, {Salam}, {Sarkar}, {Sauli}, {Schneider}, {Scholberg}, {Schwartz},
  {Scott}, {Sharma}, {Sharpe}, {Shutt}, {Silari}, {Sj{\"o}strand }, {Skands},
  {Skwarnicki}, {Smith}, {Smoot}, {Spanier}, {Spieler}, {Spiering}, {Stahl},
  {Stone}, {Sumiyoshi}, {Syphers}, {Terashi}, {Terning}, {Thoma}, {Thorne},
  {Tiator}, {Titov}, {Tkachenko}, {T{\"o}rnqvist}, {Tovey}, {Valencia}, {Van de
  Water}, {Varelas}, {Venanzoni}, {Verde}, {Vincter}, {Vogel}, {Vogt},
  {Wakely}, {Walkowiak}, {Walter}, {Wands}, {Ward}, {Wascko}, {Weiglein},
  {Weinberg}, {Weinberg}, {White}, {Wiencke}, {Willocq}, {Wohl}, {Womersley},
  {Woody}, {Workman}, {Yao}, {Zeller}, {Zenin}, {Zhu}, {Zhu}, {Zimmermann},
  {Zyla}, {Anderson}, {Fuller}, {Lugovsky}, {Schaffner}, \& {Particle Data
  Group}}]{tanabashi:18a}
{Tanabashi}, M., {Hagiwara}, K., {Hikasa}, K., {et~al.} 2018, \prd, 98, 030001,
  \dodoi{10.1103/PhysRevD.98.030001}

\bibitem[{{Taylor} {et~al.}(2006){Taylor}, {Gugliucci}, {Fabian}, {Sanders},
  {Gentile}, \& {Allen}}]{taylor:06a}
{Taylor}, G.~B., {Gugliucci}, N.~E., {Fabian}, A.~C., {et~al.} 2006, \mnras,
  368, 1500, \dodoi{10.1111/j.1365-2966.2006.10244.x}

\bibitem[{{Tiwari}(2012)}]{tiwari:12a}
{Tiwari}, P. 2012, \prd, 86, 115025, \dodoi{10.1103/PhysRevD.86.115025}

\bibitem[{{Vacca} {et~al.}(2012){Vacca}, {Murgia}, {Govoni}, {Feretti},
  {Giovannini}, {Perley}, \& {Taylor}}]{vacca:12a}
{Vacca}, V., {Murgia}, M., {Govoni}, F., {et~al.} 2012, \aap, 540, A38,
  \dodoi{10.1051/0004-6361/201116622}

\bibitem[{{Vysotsskii} {et~al.}(1978){Vysotsskii}, {Zel'dovich}, {Khlopov}, \&
  {Chechetkin}}]{vysotsskii:78a}
{Vysotsskii}, M.~I., {Zel'dovich}, Y.~B., {Khlopov}, M.~Y., \& {Chechetkin},
  V.~M. 1978, Soviet Journal of Experimental and Theoretical Physics Letters,
  27, 502

\bibitem[{{Weinberg}(1978)}]{weinberg:78a}
{Weinberg}, S. 1978, Physical Review Letters, 40, 223,
  \dodoi{10.1103/PhysRevLett.40.223}

\bibitem[{{Wilczek}(1978)}]{wilczek:78a}
{Wilczek}, F. 1978, Physical Review Letters, 40, 279,
  \dodoi{10.1103/PhysRevLett.40.279}

\bibitem[{{Wouters} \& {Brun}(2013)}]{wouters:13a}
{Wouters}, D., \& {Brun}, P. 2013, \apj, 772, 44,
  \dodoi{10.1088/0004-637X/772/1/44}

\end{thebibliography}

\end{document}